\documentclass[journal]{IEEEtran}

\IEEEoverridecommandlockouts                              

\usepackage{changes}
\usepackage{authblk}
\usepackage{amsfonts}
\usepackage{amsthm}
\usepackage{amsmath,amssymb}       
\usepackage{hyperref,cite}
\usepackage{cleveref}
\usepackage{color}
\usepackage{algorithm}
\usepackage{algorithmic}
\let\labelindent\relax
\usepackage{enumitem}
\usepackage{float}
\usepackage{graphicx}
\usepackage{subfigure}
\usepackage{booktabs}
\usepackage{threeparttable}
\usepackage{hhline}
\usepackage{caption}
\usepackage{scalerel}
\usepackage{ulem}
\usepackage{tikz}

\newtheorem{theorem}{Theorem}

\newtheorem{remark}{Remark}
\newtheorem{lemma}{Lemma}
\newtheorem{corollary}{Corollary}
 
\newcommand{\diag}{\mathop{\rm diag}\nolimits}

\newcommand{\ba}[1]{\begin{array}{#1}}
\newcommand{\ea}{\end{array}}

\newcommand*{\pdot}{\mathbin{\scalerel*{\boldsymbol\odot}{\circ}}}
\newcommand{\tikzxmark}{%
\tikz[scale=0.23] {
    \draw[line width=0.7,line cap=round] (0,0) to [bend left=6] (1,1);
    \draw[line width=0.7,line cap=round] (0.2,0.95) to [bend right=3] (0.8,0.05);
}}
\newcommand{\tikzcmark}{%
\tikz[scale=0.23] {
    \draw[line width=0.7,line cap=round] (0.25,0) to [bend left=10] (1,1);
    \draw[line width=0.8,line cap=round] (0,0.35) to [bend right=1] (0.23,0);
}}

\begin{document}
\title{\LARGE \bf
A Parallel Vector-form $LDL^\top$ Decomposition for Accelerating Execution-time-certified $\ell_1$-penalty Soft-constrained MPC}
\author{Liang Wu$^{1,*}$, Liwei Zhou$^{2}$, Richard D. Braatz$^{1}$, Fellow, IEEE%
\thanks{$^{1}$Massachusetts Institute of Technology, Cambridge, MA 02139, USA. $^{2}$Peking University, Haidian District, Beijing 100871, China. $^{*}$Corresponding author, {\tt\small liangwu@mit.edu}.}
}
\maketitle

\begin{abstract}
Handling possible infeasibility and providing an execution time certificate are two pressing requirements of real-time Model Predictive Control (MPC). To meet these two requirements simultaneously, this paper proposes an $\ell_1$-penalty soft-constrained MPC formulation that is globally feasible and solvable with an execution time certificate using our proposed algorithm. This paper proves for the first time that $\ell_1$-penalty soft-constrained MPC problems can be equivalently transformed into a box-constrained quadratic programming (Box-QP) and then our previous execution-time-certified algorithm \cite{wu2023direct} (only limited to Box-QP) can be applied. However, our previous Box-QP algorithm \cite{wu2023direct}, which provides a theoretical execution-time certificate, is conservative in its iteration analysis, thus sacrificing computation efficiency. To this end, this paper proposes a novel $LDL^\top$ decomposition for the first time, to accelerate the computation of Newton step at each iteration. The speedup of our $LDL^\top$ decomposition comes from two-fold: \textit{i)} exploitation of the fact that the number of inequality constraints is generally larger than the number of variables in condensed MPC formulations, \textit{ii)} vectorized and parallel implementation based on based on its vector-wise operations, instead of element-wise operations of previous decomposition methods.
Numerical experiments demonstrate great speedups of the proposed $LDL^\top$ decomposition (even up to 1000-fold, compared to the standard Choleksky method), which thus helps our solver achieve comparable computation performance to the state-of-the-art solvers such as IPOPT and OSQP. Code is available at \url{https://github.com/liangwu2019/L1-penalty-QP}.

\end{abstract}


\section{Introduction}
Model Predictive Control (MPC) has been successful in numerous industrial applications such as in the humanoid robot Atlas from Boston Dynamics. Except for simple problems that can compute an offline explicit form of MPC control law \cite{bemporad2002explicit}, MPC usually requires an online Quadratic Programming (QP) solver for computing the control signals.

Deploying an online QP solver related to MPC problems in real-time production environments such as embedded microcontrollers can lead to  \textit{i)} control inputs failing to return on time, or \textit{ii)} control inputs being empty upon return. Both scenarios will render the process open loop, thus resulting in safety concerns, particularly in safety-critical process systems.

Preventing the former requires a certification that the execution time of the adopted QP solver will never be greater than the sampling time. In other words, the adopted QP solver must return the solution of the current MPC problem before the next MPC problem arrives. This challenge, known as providing a theoretical execution-time certificate for real-time MPC problems, has garnered considerable interest in recent years and remains an active area of research \cite{richter2011computational,bemporad2012simple, giselsson2012execution, cimini2017exact, cimini2019complexity, arnstrom2019exact, arnstrom2020complexity, arnstrom2021unifying, okawa2021linear, wu2023direct, wu2024time, wu2024execution}. 

The latter arises from disturbances or modeling errors driving the system state into a region where the MPC problem becomes infeasible. To prevent such MPC failures, a common approach is to use soft-constraint MPC schemes that introduce slack variables to soften the constraints and add a penalty term of corresponding slack variables into the MPC cost function. Soft-constrained MPC formulations usually make solving the corresponding QP more computationally expensive and complicated. For example, \cite{kerrigan2000soft} introduced the $\ell_1$-penalty into soft-constrained MPC formulations and claimed that the $\ell_1$-penalty enjoys \textit{exact recovery} (that is, exactly recovers the same solution when the original MPC problem is feasible) without requiring the penalty parameter to be infinite, unlike the $\ell_2$-penalty. However, the $\ell_1$-penalty leads to a non-smooth QP \cite{kerrigan2000soft}, which is usually more complicated to solve.

\subsection{Related work}
Both average and worst-case (maximum) execution times are of interest, as MPC involves a sequence of QPs with varying problem data. Most studies endeavors focus on developing algorithms that possess a faster average execution time (see \cite{wang2009fast,ferreau2014qpoases, stellato2020osqp, wu2023simple, wu2023construction}). However, for deploying real-time MPC in embedded systems, the worst-case execution time is far more critical than the average execution time. This is because the sampling time is chosen to be larger than the worst-case execution time, not its average. Moreover, the worst-case execution time should be obtained from theoretical analysis rather than from the numerical experiments approach. First, it can not provide a guarantee and is numerically expensive as it tries to cover all scenarios to find the approximate worst-case execution time.

Previous works \cite{richter2011computational,bemporad2012simple, giselsson2012execution, cimini2017exact, cimini2019complexity, arnstrom2019exact, arnstrom2020complexity, arnstrom2021unifying, okawa2021linear, wu2023direct, wu2024time, wu2024execution} on providing execution-time certificates assume that the adopted computation platform performs a fixed number of floating-point operations ([flops]) in constant time units, and then \\
$$
\textrm{execution time} = \frac{\textrm{total [flops] required by an algorithm}}{\textrm{[flops] processed per second}}~[\text{s}].
$$
Calculating the total [flops] is equivalent to analyzing the number of iterations if each iteration performs fixed [flops]. For example, the backtracking line-search operation, which is widely used in most interior-point solvers, does not have a fixed [flops]. Therefore, all these works \cite{richter2011computational,bemporad2012simple, giselsson2012execution, cimini2017exact, cimini2019complexity, arnstrom2019exact, arnstrom2020complexity, arnstrom2021unifying, okawa2021linear, wu2023direct, wu2024time, wu2024execution} analyze the worst-case iteration complexity of their proposed algorithms to provide an execution time certificate for MPC applications. References \cite{giselsson2012execution, richter2011computational, bemporad2012simple} and \cite{cimini2017exact, cimini2019complexity, arnstrom2019exact, arnstrom2020complexity, arnstrom2021unifying, okawa2021linear} adopt first-order methods and active-set based methods to provide worst-case iteration complexity analysis, respectively. However, their worst-case iteration complexity results are not only complex but are \textit{data-dependent}, only allowing certification of the execution time of linear MPC. There is no extended work based on them that certifies the MPC execution time of nonlinear systems. For example, the well-known Real-Time Iteration (RTI ) scheme \cite{gros2020linear} can formulate nonlinear MPC problems as QPs but with time-varying problem data, including the Hessian matrix.

Our previous work \cite{wu2023direct} was the first to propose a box-constrained QP (Box-QP) algorithm with \textit{exact} iteration complexity $\mathcal{N}=\!\left\lceil\frac{\log(\frac{2n}{\epsilon})}{-2\log(\frac{\sqrt{2n}}{\sqrt{2n}+\sqrt{2}-1})}\right\rceil\! + 1$, which is \textit{data-independent} (with cost only dependent on dimension) and thus able to certify the execution time of input-constrained nonlinear MPC problems via the RTI scheme \cite{wu2024execution} and Koopman operator \cite{wu2024time}. Our \textit{simple-to-calculate} \textit{dimension-dependent} iteration complexity is also valuable for the MPC design phase because it allows the selection of the appropriate model complexity, prediction horizon length, and sampling time to ensure compliance with the execution-time certificate requirement, instead of performing extensive simulations or heavy calibration work \cite{forgione2020efficient}. 

However, our Box-QP algorithm was limited to input-constrained MPC problems and did not apply to general MPC problems with both input and state constraints. 

\subsection{Contribution}
This paper adopts an $\ell_1$-penalty soft-constrained MPC formulation to handle the potential infeasibility of real-time MPC problems with both input and state constraints. Then, it is proved for the first time that the $\ell_1$-penalty soft-constrained MPC problems can be equivalently transformed into a Box-QP which allows our previous execution-time-certified Box-QP algorithm \cite{wu2023direct} to be applied. Moreover, to improve computation efficiency, this paper proposes, for the first time, a novel parallel vector-form $LDL^\top$ decomposition. It exploits the problem's structure and leverages the vectorized and parallel implementation, resulting in significant speedups in the computation of the Newton step at each iteration. That allows our proposed algorithm to achieve comparable computation performance to state-of-the-art solvers such as IPOPT and OSQP while still preserving an execution-time certificate.

The proposed algorithm is the first to be able to handle possible infeasibility and provide an execution time certificate of real-time general MPC problems with both input and state constraints.

\subsection{Notations}
The vector of all ones is denoted by $e=(1,\cdots{},1)^\top$. $\left\lceil x\right\rceil$ maps $x$ to the least integer greater than or equal to $x$. For $z,y\in\mathbb{R}^n$, let $\mathrm{col}(z,y)=[z^{\top},y^{\top}]^{\top}$ and $\max(z,y)=(\max(z_1,y_1),\max(z_2,y_2),\cdots{},\max(z_n,y_n))^\top$. $\mathbb{R}^n$ denotes the space of $n$-dimensional real vectors, $\mathbb{R}^n_{++}$ is the set of all positive vectors of $\mathbb{R}^n$. 
For a vector $z\in\mathbb{R}^n$, its Euclidean norm is $\|z\|=\sqrt{z_1^2+z_2^2+\cdots+z_n^2}$, $\|z\|_1=\sum_{i=1}^{n}|z_i|$, $\|z\|_{\infty}=\max_i |z_i|$, $\diag(z):\mathbb{R}^n\rightarrow\mathbb{R}^{n\times n}$ maps an vector $z$ to its corresponding diagonal matrix, $z^2 = (z_1^2,z_2^2,\cdots{},z_n^2)^\top$, $\sqrt{z}=\left(\sqrt{z_1}, \sqrt{z_2},\cdots{},\sqrt{z_n}\right)^{\!\top}$ and $z^{-1}=(z_1^{-1},z_2^{-1},\cdots{},z_n^{-1})^{\top}$. Given two arbitrary vectors $z,y \in\mathbb{R}^n_{++}$, their Hadamard product is $z\pdot y = (z_1y_1,z_2y_2,\cdots{},z_ny_n)^\top$, $z\pdot \big(y^{-1}\big)=\big(\frac{z_1}{y_1},\frac{z_2}{y_2},\cdots{},\frac{z_n}{y_n}
\big)^{\!\top}$.

\section{Problem formulation}
We consider linear MPC problems for tracking,
\begin{equation}\label{problem_MPC_tracking}
    \begin{aligned}
        \min~&\frac{1}{2}\sum_{k=0}^{N_p-1}\left\|W_y\! \left(Cx_{k+1}-r(t)\right)\right\|_{2}^{2} + \left\|W_{\Delta u}  \Delta u_{k}\right\|_{2}^{2} \\
        \mathrm{s.t.}~& \forall k=0, \ldots, N_p-1,\\
        &x_{k+1}=A x_{k}+B u_{k},\\
        & u_{k} = u_{k-1} + \Delta u_{k},\\
        & E_x x_{k+1} + E_u u_k + E_{\Delta u} \Delta u_{k}\leq f_k,\\
        &x_0 = x(t), u_{-1} = u(t-1),
    \end{aligned}
\end{equation}
where $x_k\in\mathbb{R}^{n_x}$ denotes the states, $u_k\in\mathbb{R}^{n_u}$ denotes the control inputs. The control objective is to minimize the tracking errors between the output $Cx_{k+1}$ and the reference signal $r(t)$, and the control input increments $\Delta u_k$, along the prediction horizon $N_p$. $W_y\succ0$ and $W_{\Delta u}\succ0$ are the weights for output tracking error and control input increments, respectively. The control objective sometimes adds the extra penalty $\|W_u(u_k-u_{\text{ref}}(t)\|^2_2$ to track control input references $u_{\text{ref}}(t)$. $x(t)$ and $u(t-1)$ denote the current feedback states and previous control inputs, respectively. $E_x x_{k+1} + E_u u_k + E_{\Delta u} \Delta u_{k}\leq f_k$ denotes the general constraints formulation such as the box constraints $u_{\min } \leq u_{k} \leq u_{\max}, x_{\min} \leq x_{k+1} \leq x_{\max},\Delta u_{\min } \leq \Delta u_{k} \leq \Delta u_{\max}$ or the terminal constraints $x_{N_p}\in\mathcal{X}_{N_p}$.

Define $y\triangleq\mathrm{col}(\Delta u_0, \Delta u_1, \cdots{}, \Delta u_{N_p-1})\in\mathbb{R}^{m}$ ($m=N_pn_u$) and $\Bar{x}(t)\triangleq\mathrm{col}(x(t),u(t-1),r(t))$. Then, by eliminating all equality constraints (the condensing construction approach, see \cite{jerez2011condensed}), Problem \eqref{problem_MPC_tracking} can be formulated as the general QP:
\begin{equation}\label{problem_QP}
\begin{aligned}
\min_{y}~& \frac{1}{2}y^\top Q y+y^\top (F\Bar{x}(t))\\
\mathrm{s.t.}~& Gy\leq g + S\Bar{x}(t)
\end{aligned}   
\end{equation}
where the problem data such as the symmetric positive definite matrix $Q\in\mathbb{R}^{m\times m}\succ0$, the matrix $G\in\mathbb{R}^{n\times m}$ ($n$ denotes the total number of inequality constraints), the matrices $F$ and $S$, and the vector $g\in\mathbb{R}^n$ can be offline computed except the optimization data $\Bar{x}(t)$ are time-varying. 
\begin{remark}\label{remark_n_geq_m}
The resulting QP \eqref{problem_QP} from condensing MPC formulations, usually has a larger number of inequality constraints than the number of decision variables, that is, $n\geq m$, or even $n\gg m$ when the number of state constraints is larger than the control inputs.    
\end{remark}

The QP (\ref{problem_QP}) may be infeasible when a disturbance or modeling errors (from linearization approximation) drive the system into a region that the set $Gy\leq g+S\bar{x}(t)$ of the QP (\ref{problem_QP}) is empty. To ensure  feasibility, the $\ell_1$ penalty approach is adopted to soften the QP (\ref{problem_QP}). The $\ell_1$-penalty method reformulates (\ref{problem_QP}) as the unconstrained non-smooth penalty minimization,
\begin{equation}\label{problem_QP_to_penaltyQP}
\begin{aligned}
    \min_y&\quad\frac{1}{2}y^\top Qy+y^\top (F\bar{x}(t))\\
    &~+\rho^\top\|\max(0,Gy-g-S\bar{x}(t))\|_1
\end{aligned}
\end{equation}
where $\rho\in\mathbb{R}_{++}^n$ denotes the penalty weight vector. Compared to the most frequently used $\ell_2$ quadratic penalty method, which requires $\rho$ to be very large (see \cite{saraf2017fast}), the $\ell_1$-penalty method does not require large $\rho$. Because a finite value is enough when the penalty weight $\rho$ is larger than the corresponding optimal Lagrange multiplier value according to \cite[Thm. 14.3.1]{fletcher2000practical}, $\ell_1$-penalty QP \eqref{problem_QP_to_penaltyQP} can exactly recover the same solution when the original QP \eqref{problem_QP} is feasible. The argument that the $\ell_1$-penalty method is better can also be explained from the physical meaning, in which the $\ell_1$-penalty encourages sparsity in constraint violation and thus encourages the solution to satisfy as many of the constraints as possible.

MPC distinguishes each constraint as being either hard or soft. The hard constraints are from the physical limits of actuators (the control inputs), which should never be violated. The soft constraints are user-specified state constraints, which can be violated by a small amount. By setting larger penalty weights for constraints with higher priorities (such as hard constraints having the largest penalty weights), we can avoid physically invalid solutions. 

\begin{lemma}\label{lemma_global_feasibility}
    Suppose that $Q$ is symmetric positive definite, the $\ell_1$-penalty non-smooth QP \eqref{problem_QP_to_penaltyQP} is globally feasible with a unique optimal solution $y^*$.
\end{lemma}
\begin{proof}
    The regularization term $\rho^\top\|\max(0, Gy - g- S\bar{x}(t)\|_1$ involves the $\ell_1$ norm, a convex function composed with the ReLu function $\max(0, Gy-g-S\bar{x}(t))$, which is a convex function composited with a convex function $Gy-g-S\bar{x}(t)$. The composition of convex functions remains convex. Thus, the overall objective function is strictly convex when $Q$ is symmetric positive definite, which guarantees the uniqueness of the optimal solution.
\end{proof}

By Lemma \ref{lemma_global_feasibility}, the $\ell_1$-penalty soft-constrained MPC from Problem \ref{problem_MPC_tracking} can be formulated to be globally feasible in closed-loop. After that, the non-smooth $\ell_1$-penalty QP \eqref{problem_QP_to_penaltyQP} is proved to be equivalent to a Box-QP.

\subsection{Transform non-smooth $\ell_1$-penalty QP to Box-QP}
By introducing slack variables 
\[
w=\max\left(0,Gy-g-S\bar{x}(t)\right)\in\mathbb{R}^n,
\]
we obtain Corollary 1.
\begin{corollary}\label{corollary_smoothQP}
    The non-smooth $\ell_1$-penalty QP \eqref{problem_QP_to_penaltyQP} is equivalent to the smooth convex QP,
\begin{equation}\label{problem_penaltyQP_to_smoothQP}
    \begin{aligned}
        \min_{y,w}~&\frac{1}{2}y^\top Q y+y^\top (F\bar{x}(t)) + \rho^\top w\\
        \mathrm{s.t}~&w\geq0\\
        ~&w\geq Gy-g-S\bar{x}(t)
    \end{aligned}    
\end{equation}
and it has a unique optimal solution $\mathrm{col}(y^*, w^*)$, in which $y^*$ is the same as the unique optimal solution of the non-smooth $\ell_1$-penalty QP \eqref{problem_QP_to_penaltyQP}.
\end{corollary}
\begin{proof}
The introduction of vector $w\in\mathbb{R}$ reduces the $\ell_1$ term to an affine linear term,
\[
\rho^\top\,\|\!\max(0,G(t)y-g(t)-S(t)x(t))\|_1=\rho^\top w,
\]
with two additional constraints
\[
w\geq0, ~w\geq Gy-g-S\bar{x}(t).
\]
Thus, the non-smooth $\ell_1$-penalty QP \eqref{problem_QP_to_penaltyQP} is equivalent to the smooth convex QP \eqref{problem_penaltyQP_to_smoothQP}. Since \eqref{problem_penaltyQP_to_smoothQP} inherits the same optimal solution $y^*$ of \eqref{problem_QP_to_penaltyQP}, which exists and is unique by Lemma \ref{lemma_global_feasibility}, $w^*=\max(0,Gx-g-S\bar{x}(t))$ exists and is unique. Thus, the solution of the smooth convex QP \eqref{problem_penaltyQP_to_smoothQP} exists and is unique, despite the loss of strict convexity (the Hessian matrix of \eqref{problem_penaltyQP_to_smoothQP} is $\diag(Q,0)\succeq0$).
\end{proof}

The Lagrangian function $\mathcal{L}(y,w,\Tilde{z},\hat{z})$ of the smooth convex QP \eqref{problem_penaltyQP_to_smoothQP} is given by
\[
\begin{aligned}
    \mathcal{L}(y,w,\Tilde{z},\hat{z}) &= \frac{1}{2}y^\top Q y+y^\top (F\bar{x}(t)) + \rho^\top w \\
    & \quad - \Tilde{z}^\top w - \hat{z}^\top (w-Gy+g+S\bar{x}(t))
\end{aligned}
\]
where $\Tilde{z},\hat{z}\in\mathbb{R}^n$ are the vector of Lagrange multipliers associated with the inequalities in (\ref{problem_penaltyQP_to_smoothQP}). By introducing the slack variable $\delta\triangleq  w-Gy+g+S\bar{x}(t)$, the Karush–Kuhn–Tucker (KKT) conditions of the smooth convex QP \eqref{problem_penaltyQP_to_smoothQP} are
\begin{subequations}\label{eqn_KKT_QP}
\begin{align}
        &Qy + F\bar{x}(t)+G^\top \hat{z} = 0\label{eqn_KKT_QP_a}\\
        &\delta=  w-Gy+g+S\bar{x}(t)\label{eqn_KKT_QP_b}\\
        &\rho - \Tilde{z} - \hat{z}=0\label{eqn_KKT_QP_c}\\
        &w\geq0,~\Tilde{z}\geq0,~w_i\Tilde{z}_i=0,~\forall i=1,\cdots{},n\label{eqn_KKT_QP_d}\\
        &\delta\geq0,~\hat{z}\geq0,~\delta_i\hat{z}_i=0,~\forall i=1,\cdots{},n\label{eqn_KKT_QP_e}
\end{align}
\end{subequations}

\begin{theorem}\label{theorem_1}
    Suppose that $Q$ is symmetric positive definite, the non-smooth $\ell_1$-penalty QP \eqref{problem_QP_to_penaltyQP} is equivalent to the convex Box-QP
\begin{equation}\label{problem_box_QP_0}
    \begin{aligned}
        \min_{\hat{z}}~&\frac{1}{2}\hat{z}^\top M\hat{z}+\hat{z}^\top d\\
        \mathrm{s.t.}~&0\leq\hat{z}\leq\rho
    \end{aligned}
\end{equation}
where $M=GQ^{-1}G^\top$ and $d=GQ^{-1}F\bar{x}(t)+g+S\bar{x}(t)$. Furthermore, Box-QP (\ref{problem_box_QP_0}) has a unique optimal solution $\hat{z}^*$, which can be used to recover the unique optimal solution $y^*$ of the non-smooth $\ell_1$-penalty QP  \eqref{problem_QP_to_penaltyQP} by 
\begin{equation}\label{eqn_optimal_solution_relationship}
    y^* = -Q^{-1}(F\bar{x}(t)+G^\top \hat{z}^*).
\end{equation}
\end{theorem}
\begin{proof}
By Corollary \ref{corollary_smoothQP}, the smooth convex QP \eqref{problem_penaltyQP_to_smoothQP} has a unique optimal solution, which proves that the KKT conditions \eqref{eqn_KKT_QP} hold for a unique solution \cite{boyd2004convex}.

For reducing the KKT conditions \eqref{eqn_KKT_QP} into a compact condition, \eqref{eqn_KKT_QP_a} can be eliminated by using $y = -Q^{-1}(F\bar{x}(t)+G^\top \hat{z})$ as $Q$ is symmetric positive definite. Then, substitution into \eqref{eqn_KKT_QP_b}  eliminates $y$; \eqref{eqn_KKT_QP_c} can also be eliminated by using $\tilde{z}=\rho-\hat{z}$. This allows the KKT conditions \eqref{eqn_KKT_QP} to be reduced to
\begin{equation}\label{eqn_KKT_reformulated}
    \begin{aligned}
        &M\hat{z}+d-\delta+w=0\\
        &\delta\geq0,~\hat{z}\geq0,~\delta_i\hat{z}_i=0,~\forall i=1,\cdots{},n\\
        &w\geq0,~\rho-\hat{z}\geq0, ~w_i(\rho_i -\hat{z}_i)=0,~\forall i=1,\cdots{},n
    \end{aligned}
    \end{equation}
where $M=GQ^{-1}G^\top$ and $d=GQ^{-1}F\bar{x}(t)+g+S\bar{x}(t)$. Thus, the new KKT condition \eqref{eqn_KKT_reformulated} also has a unique solution. Eq.\ \eqref{eqn_KKT_reformulated} is the KKT condition of Box-QP \eqref{problem_box_QP_0}, therefore Box-QP (\ref{problem_box_QP_0}) has a unique optimal solution $\hat{z}^*$ that uniquely corresponds to $y^*=-Q^{-1}(F\bar{x}(t)+G^\top \hat{z}^*)$.
\end{proof}

\subsection{Scaling the Box-QP}
Thanks to Theorem \ref{theorem_1}, we turn to solve Box-QP \eqref{problem_box_QP_0}, which can be solved by our previous algorithm with an execution-time certificate \cite{wu2023direct}. Before that, we scale the Box-QP (\ref{problem_box_QP_0}) to the box constraint $[-e,e]$. Applying the coordinate transformation
\[
z = \diag\!\Big(\frac{2}{\rho}\Big)\hat{z}-e
\]
results in the equivalent scaled Box-QP,
\begin{subequations}\label{problem_scaled_box_QP}
\begin{align}
    \min_{z}~&\frac{1}{2}z^\top H z + z^\top h\label{problem_scaled_box_QP_objective}\\
    \mathrm{s.t.}~& -e \leq z \leq e \label{problem_scaled_box_QP_constraint}
\end{align}
\end{subequations}
where $H=\diag(\rho)M\diag(\rho)$ and $h=\diag(\rho)(M\rho +2d)$. Based on the equivalence and by Theorem \ref{theorem_1}, we have Corollary 2.
\begin{corollary}
    The unique optimal solution $y^*$ of the non-smooth $\ell_1$-penalty QP \ref{problem_QP_to_penaltyQP} can be recovered by the unique optimal solution $z^*$ of the scaled Box-QP (\ref{problem_scaled_box_QP}) with
    \begin{equation}
        y^* = -Q^{-1}\!\left(F\bar{x}(t)+\frac{1}{2}G^\top(\rho z^*+\rho)\right).
    \end{equation}
\end{corollary}

\section{Execution-time-certified QP algorithm}
The Karush–Kuhn–Tucker (KKT) condition of Box-QP \eqref{problem_scaled_box_QP} is 
\begin{subequations}\label{eqn_KKT}
\begin{align}
    Hz + h + \gamma - \theta = 0\label{eqn_KKT_a}\\
    z + \phi - e=0\label{eqn_KKT_b}\\
    z - \psi + e=0\label{eqn_KKT_c}\\
    \gamma \pdot \phi = 0\label{eqn_KKT_d}\\
    \theta \pdot \psi = 0\label{eqn_KKT_e}\\
    (\gamma,\theta,\phi,\psi)\geq0
\end{align}
\end{subequations}
where $\gamma \in\mathbb{R}^n$, $\theta \in\mathbb{R}^n$, $\phi \in\mathbb{R}^n$, and $\psi \in\mathbb{R}^n$ denote the Lagrangian variable and the slack variable of the lower bound and upper bound, respectively. 

Our previous Box-QP algorithm \cite{wu2023direct} is based on the path-following Interior-point method (IPM), which introduces a positive parameter $\tau>0$ and an element-wise \textit{square root} function to replace \eqref{eqn_KKT_d} and \eqref{eqn_KKT_e} by
\begin{subequations}\label{eqn_KKT_tau}
\begin{align}
    \sqrt{\gamma \pdot \phi} = \tau e\label{eqn_KKT_tau_d}\\
    \sqrt{\theta \pdot \psi} = \tau e\label{eqn_KKT_tau_e}
\end{align}
\end{subequations}
There exists a unique solution $(z_{\tau},\gamma_{\tau},\theta_{\tau},\phi_{\tau},\psi_{\tau})$ for the nonlinear system (\eqref{eqn_KKT_a}--\eqref{eqn_KKT_c}, \eqref{eqn_KKT_tau_d}--\eqref{eqn_KKT_tau_e}) and, as $\tau$ approaches 0,  $(z_{\tau},\gamma_{\tau},\theta_{\tau},\phi_{\tau},\psi_{\tau})$ goes to the solution of \eqref{eqn_KKT}. At each iteration, the search direction is determined by using Newton's method which involves linearizing the nonlinear system (\eqref{eqn_KKT_a}--\eqref{eqn_KKT_c}, \eqref{eqn_KKT_tau_d}--\eqref{eqn_KKT_tau_e}) and solving the linearized Newton system. As for the step size, to ensure iterates satisfy \eqref{eqn_KKT_e} and converge, our previous work \cite{wu2023direct} proves that the step size can be 1 (full-Newton step) to make all iterates lie in the strictly feasible set
\[
    \mathcal{F}^0=\{(z,\gamma,\theta,\phi,\psi)|\,\text{satisfying}\, \eqref{eqn_KKT_a}\textrm{--}\eqref{eqn_KKT_c},(\gamma,\theta,\phi,\psi)>0\}
\]
and achieve the \textit{exact} iteration complexity, if we choose the appropriate initialization strategy for $(z^0,\gamma^0,\theta^0,\phi^0,\psi^0)$ and decreasing strategy for $\tau$.

\subsection{Strictly feasible initial point}
Box-QP allows us to, at no computation cost, find a strictly feasible initial point belonging to $\mathcal{F}^0$ such as 
\begin{equation*}
z^0 = 0, \gamma^0= \|h\|_\infty - \tfrac{1}{2}h,\theta^0 =\|h\|_\infty + \tfrac{1}{2}h,\phi^0 = e,\psi^0 = e,
\end{equation*}
where $\|h\|_\infty=\max \{ |h_1|,|h_2|,\cdots{},|h_{n}|\}$, but this condition is not sufficient because achieving the \textit{exact} iteration complexity requires that all iterates (including the strictly feasible initial point) be located in a narrow neighborhood of the central path. To ensure this, our previous work \cite{wu2023direct} scales the objective \eqref{problem_scaled_box_QP_objective} as 
\[
\min_z \tfrac{1}{2} z^\top \!\!\left(\frac{2\lambda}{\|h\|_\infty}H\right) \!z + z^\top\! \!\left(\frac{2\lambda}{\|h\|_\infty}h\right)\!,
\]
where $\lambda>0$, which has no effect on the optimal solution for $h\neq 0$ (for the case $h=0$, the optimal solution of \eqref{problem_scaled_box_QP} is $z^*=0$). Defining $\tilde{H} \triangleq \frac{1}{\|h\|_\infty}H$ and $\tilde{h}\triangleq\frac{1}{\|h\|_\infty}h$, \cite[Remark 1]{wu2023direct} suggests the initialization strategy,
\begin{equation}\label{eqn_initialization_stragegy}
 z^0 = 0,\ \gamma^0= 1 - \lambda \tilde{h},\ \theta^0 = 1 + \lambda \tilde{h},\ \phi^0 = e, \ \psi^0 = e,
\end{equation}
to force the initial point to lie within the narrow neighborhood of the central path by setting $\lambda=\frac{1}{\sqrt{n+1}}$ according to \cite[Lemma 4]{wu2023direct}.

\subsection{Newton direction}
After adopting the above scaling and initialization strategy, suppose that, for the current iterate $(z,\gamma,\theta,\phi,\psi)\in\mathcal{F}^0$, a direction $(\Delta z,\Delta\gamma, \Delta\theta, \Delta\phi \Delta\psi)$ is computed by solving the linearized KKT equation around the current iterate as
 \[
 \begin{aligned}
     2\lambda\tilde{H}\Delta z +  \Delta \gamma - \Delta \theta &= 0\\
     \Delta z + \Delta\phi &= 0\\
     \Delta z - \Delta\psi &= 0\\
     \left(\!\sqrt{\phi\pdot\gamma^{-1}}\right)\!\pdot \Delta\gamma + \!\left(\!\sqrt{\gamma\pdot\phi^{-1}}\right)\!\pdot\Delta\phi &= 2(\tau e- \!\sqrt{\gamma\pdot\phi})\\
     \left(\!\sqrt{\psi\pdot\theta^{-1}}\right)\!\pdot\Delta\theta + \!\left(\!\sqrt{\theta\pdot\psi^{-1}}\right)\!\pdot\Delta\psi &= 2(\tau e - \!\sqrt{\theta\pdot\psi})
 \end{aligned}
 \]
Then, by setting 
\begin{subequations}\label{eqn_Delta_gamma_theta_phi_psi}
    \begin{align}
        &\Delta \gamma=\!\left(\gamma\pdot\phi^{-1}\right)\!\pdot\Delta z+2\!\left(\tau\sqrt{\gamma\pdot\phi^{-1}}-\gamma\right)\!,\\
        &\Delta \theta=-(\theta\pdot\psi^{-1})\pdot\Delta z+2\!\left(\tau\sqrt{\theta\pdot\psi^{-1}}-\theta\right)\!,\\
        &\Delta\phi = - \Delta z,\\
        &\Delta\psi = \Delta z,
    \end{align}
\end{subequations}
we can obtain a more compact system of linear equations as
\begin{equation}\label{eqn_compact_linsys}
    \begin{aligned}
    \Big(2\lambda\tilde{H}+&\diag\!\Big(\frac{\gamma}{\phi}\Big) + \diag\!\Big(\frac{\theta}{\psi}\Big) \!\Big) \Delta z\\
        &=2\!\left(\tau\sqrt{\theta\pdot\psi^{-1}} - \tau\sqrt{\gamma\pdot\phi^{-1}}+ \gamma - \theta\right).
    \end{aligned}
\end{equation}

\subsection{Iteration complexity and algorithm implementation}
Our previous work \cite{wu2023direct} proved that, by choosing the appropriate decreasing rule of $\tau$, the new iterate, after taking a full-Newton step, that is, $z\leftarrow z+\Delta z, \gamma\leftarrow\gamma+\Delta\gamma, \theta\leftarrow\theta+\Delta\theta,\phi\leftarrow\phi+\Delta\phi, \psi\leftarrow\psi+\Delta\psi$, still lies in $\mathcal{F}^0$. Our previous algorithm was
\begin{equation}\label{procedure_ETC_BoxQP}
  \begin{aligned}
  &\textbf{For}~\text{iter}=1,\cdots{}, \mathcal{N}\\
    &\quad\text{(1)}~\tau \leftarrow (1-\eta)\tau;\\
    &\quad\text{(2)}~\text{solve}~\eqref{eqn_compact_linsys}~\text{for}~\Delta z;\\
    &\quad\text{(3)}~\text{calculate}~(\Delta\gamma, \Delta\theta, \Delta\phi,\Delta\psi)~\text{from}~ \eqref{eqn_Delta_gamma_theta_phi_psi};\\
    &\quad\text{(4)}~z\leftarrow z+\Delta z, \gamma\leftarrow\gamma+\Delta\gamma, \theta\leftarrow\theta+\Delta\theta,\\
    &\quad\quad~\,\phi\leftarrow\phi+\Delta\phi, \psi\leftarrow\psi+\Delta\psi;
\end{aligned}  
\end{equation}
where the value of $\mathcal{N}$ in guaranteeing convergence is illustrated by Lemma 2.
\begin{lemma}[See Theorem 2, \cite{wu2023direct}]\label{lemma_exact}
    Let $\eta=\frac{\sqrt{2}-1}{\sqrt{2n}+\sqrt{2}-1}$ and $\tau^0 =\frac{1}{1-\eta}$, Algorithm \eqref{procedure_ETC_BoxQP} exactly requires
    \begin{equation}\label{eqn_worst_number_iteragions}
    \mathcal{N}=\!\left\lceil\frac{\log(\frac{2n}{\epsilon})}{-2\log(\frac{\sqrt{2n}}{\sqrt{2n}+\sqrt{2}-1})}\right\rceil\! + 1
    \end{equation}
iterations, which gives that $\gamma^\top \phi+\theta^\top\psi \leq\epsilon$.
\end{lemma}

\section{Parallel Vector-form $LDL^\top$ Decomposition} 
The most expensive step in Algorithm \eqref{procedure_ETC_BoxQP} is to solve the symmetric positive definite system \eqref{eqn_compact_linsys}. As $2\lambda\Tilde{H} = \frac{2\lambda}{\|h\|_\infty}\diag(\rho)GQ^{-1}G\diag(\rho)$, we set
\[
\begin{aligned}
V &\triangleq \sqrt{\frac{2\lambda}{\|h\|_\infty}}\diag(\rho)GL_Q^{-\top}\!\in\mathbb{R}^{n\times m}, \\
D &\triangleq \diag\!\left(\frac{\gamma}{\phi}\right) + \diag\!\left(\frac{\theta}{\psi}\right),\\
p &\triangleq 2\!\left(\tau\sqrt{\theta\pdot\psi^{-1}} - \tau\sqrt{\gamma\pdot\phi^{-1}}+ \gamma - \theta\right),
\end{aligned}
\]
where $L_Q$ is the Cholesky factor of $Q=L_QL_Q^\top$. $L_Q$ is also necessary in \eqref{eqn_optimal_solution_relationship} for recovering $y^*$, and thus can be precomputed offline. Then, \eqref{eqn_compact_linsys} is reformulated as 
\begin{equation}\label{eqn_VV_D_z_u}
    (VV^\top + D)\Delta z = \!\left(\sum_{i=1}^m(v^i)(v^i)^\top+D\right)\!\Delta z = p
\end{equation}
where $v^i\in\mathbb{R}^n$ is $i$th column of $V$ and thus \eqref{eqn_VV_D_z_u} can regarded as rank-$m$ modifications of a diagonal system, the low-rank feature as stated by Remark \ref{remark_n_geq_m}. Exploiting this low-rank feature usually achieves significant speedup, but numerical stability and programmatic vectorization, are also essential for a numerical implementation.

\begin{table}[!htbp]
    \caption{Pros and cons of different decomposition methods}
    \centering
    \begin{tabular}{cccc}
    \toprule
     Decomposition  & Exploit & Numerically  & Programmatically\\
     methods &  $n\geq m$ & stable &  vectorizable\\
     \midrule
      Standard Cholesky & \tikzxmark  & \tikzcmark & \tikzxmark \\
      Sherman-Morrison & \tikzcmark & \tikzxmark  & \tikzxmark\\
      Product-form $LDL^\top$ & \tikzcmark & \tikzcmark & \tikzxmark \\
      Vector-form $LDL^\top$ & \tikzcmark & \tikzcmark  & \tikzcmark\\
     \bottomrule
    \end{tabular}
    \label{tab:pros_cons_decompositions}
\end{table}

Table \ref{tab:pros_cons_decompositions} lists the pros and cons of the standard Cholesky, Sherman-Morrison(-Woodbury), and Product-form $LDL^\top$ decomposition methods. As all have drawbacks, we propose for the first time a parallel vector-form $LDL^\top$ decomposition method that has none of these drawbacks.

The standard Cholesky decomposition is a general decomposition method for symmetric positive-definite systems that is widely used in IPMs for its numerical stability. But it can not exploit the low-rank feature $n\geq m$ and its operations are element-wise, which are not readily amenable to vectorization and parallelization and thus cannot be accelerated in parallel computing architectures.  

To exploit the low-rank feature, one alternative is to use the Sherman-Morrison-Woodbury identity\cite{max1950inverting}, 
\[
(VV^\top+D)^{-1}=D^{-1}-D^{-1}V(V^\top DV+I_m)^{-1}V^\top D^{-1}.
\]
Solving \eqref{eqn_VV_D_z_u} is equivalent to
\begin{subequations}\label{eqn_sherman_morrison}
    \begin{eqnarray}
        &&\Delta \tilde{z} \leftarrow D^{-1}p,\\
        &&\Delta \hat{z}: (V^\top D^{-1}V+I_m)\Delta \hat{z} = V^\top \Delta \tilde{z},\label{eqn_small_Cholesky}\\
        &&\Delta z \leftarrow \Delta \tilde{z} -D^{-1}V\Delta \hat{z}.
    \end{eqnarray}
\end{subequations}
where the second step \eqref{eqn_small_Cholesky} relies on the element-wise Cholesky decomposition, which has a smaller $\frac{1}{3}m^3+\frac{1}{2}m^2+\frac{1}{6}m$ [flops]. However, this approach is often numerically less stable in IPMs such as at the end stage where the duality gap approaches zero, as highlighted by \cite{fine2001efficient} with several examples. As pointed out by \cite{fine2001efficient}, the reason for its poor numerical stability is that it uses rank updates for the inverse of the matrix, instead of the matrix itself as done by the Cholesky decomposition approach.

In \cite{fine2001efficient}, a product-form $LDL^\top$ decomposition is proposed to exploit the low-rank feature and preserve the numerical stability, as follows,
\begin{equation}\label{eqn_prod_LDL}
VV^\top +D=L^1L^2\cdots L^mD^m(L^m)^\top\cdots(L^2)^\top(L^1)^\top,   
\end{equation}
where each $L^l$ is a special lower triangular matrix and $\Tilde{D}$ is a positive diagonal matrix. In \cite{fine2001efficient}, each $L^l$ consists in two vectors $\tilde{v}^l\in\mathbb{R}^m$ and $b^l\in\mathbb{R}^m$, and its special form is
\begin{eqnarray}\label{eqn_prodLDL_L}
 L^l(\tilde{v}^l, b^l) = \left[\begin{array}{@{}c@{}c@{}c@{}c@{}}
        1 &   & &  \\
\tilde{v}^l_2 b^l_1 & 1 & & \\
  \vdots  & \ddots & \ddots  &  \\
        \tilde{v}^l_n b^l_1  & \ \cdots \ & \tilde{v}^l_n b^l_{n-1} & \ 1
    \end{array}\right]    
\end{eqnarray}
where $\tilde{v}^l_i, b^l_i$ denotes the $i$-th element of vector $\tilde{v}^l$ and $b^l$, respectively. The decomposition procedure is equivalent to computing all $\tilde{v}^l$ and $b^l$ and $D^m$, whose derivation is based on the recursive rank-1 update of $LDL^\top$, see Appendix \ref{sec_appendix_prodLDL}. The product-form $LDL^\top$ achieves smaller $O(nm^2)$ [flops] than the Cholesky decomposition. The product-form $LDL^\top$ decomposition has been used in IPMs for linear programming, second-order cone programming, and convex QP, respectively see \cite{goldfarb2004product,goldfarb2005product,goldfarb2008numerically}. However,
the element-wise and sequential operations of the product-form  $LDL^\top$ decomposition approach even hinder achieving the expected speedup, let alone a further acceleration via vectorized and parallel programming.

To address all these drawbacks, we propose for the first time a parallel vector-form $LDL^\top$ decomposition method.

\subsection{Decomposition procedure of parallel vector-form $LDL^\top$}
Inspired by the derivation of the product-form $LDL^\top$ decomposition, this paper proposes a parallel vector-form  $LDL^\top$ decomposition approach for solving \eqref{eqn_VV_D_z_u}. Our approach directly factorizes $VV^\top + D$ as
\begin{equation}\label{eqn_LDL_decomposition}
    VV^\top + D = L(V,B) \Tilde{D}  L(V,B)^\top
\end{equation}
where the matrix $B\in\mathbb{R}^{m\times n}$ and positive diagonal matrix $\Tilde{D}\in\mathbb{R}^{n\times n}$ are unknown matrices to be computed. Inspired by the structure of \eqref{eqn_prodLDL_L}, $L(V, B)$ is a special lower triangular matrix whose subdiagonal elements are the inner product of two vectors from $V$ and $B$ as
\begin{equation}\label{eqn_L_V_B}
    L(V,B) = \left[\begin{array}{@{}c@{}c@{}c@{}c@{}}
        1 & & & \\
        (v^2)^{\!\top} b^1 & 1 & &\\
        \vdots & & \ddots &  \\
         (v^n)^{\!\top}  b^1 & \ \cdots \ &  (v^n)^{\!\top} b^{n-1} & \ 1
    \end{array} \right], 
\end{equation}
i.e.,
\[
L(V,B)_{ij}=\left\{\begin{array}{@{}cc}
        0 & \text{for}~i<j \\
         1 & \text{for}~i=j\\
        (v^i)^{\!\top} b^j &\text{for}~ i>j
    \end{array} \right.
\]
where $v^i\in\mathbb{R}^m$ denotes the $i$-th column vector of $V^\top\in\mathbb{R}^{m\times n}$ and $b^i\in\mathbb{R}^m$ denotes the $i$-th column vector of B, which is why we call this a vector-form $LDL^\top$ decomposition. The vectors $v^1$ and $b^n$ do not appear in \eqref{eqn_L_V_B}, as $v^1$ is used to calculate the first element of $\tilde{D}$ and $b^n$ is not necessary (see \eqref{eqn_LDL_b_j}).

Deriving the computation of $B$ and $D$ is based on the element-wise notation of \eqref{eqn_LDL_decomposition} as follows,
\begin{equation}\label{eqn_v_i_v_i}
     (v^i)^\top  v^i + D_{i,i} = \Tilde{D}_{i,i} + \sum_{k=1}^{i-1}\tilde{D}_{k,k}\left( (v^i)^\top b^k \right) \left( (b^k)^\top v^i\right)
\end{equation}
and, for all $j<i$,
\begin{equation}\label{eqn_v_i_v_j}
     (v^i)^\top v^j = \tilde{D}_{j,j} (v^i)^\top b^j+ \sum_{k=1}^{j-1}\Tilde{D}_{k,k}\left( (v^i)^\top b^k \right) \left( (b^k)^\top v^j \right) \
\end{equation}
Then define a sequence of auxiliary matrices $M_j$ by
\[
M^j \triangleq I_m - \sum_{k=1}^{j-1}\tilde{D}_{k,k}b^k(b^k)^\top
\]
Clearly, $M^j$ satisfies the recurrence,
\begin{equation}
    M^1 = I_m,~ M^{j+1} = M^j-\tilde{D}_{j,j} b^j (b^j)^\top.
\end{equation}
After that, we can rewrite \eqref{eqn_v_i_v_i} and \eqref{eqn_v_i_v_j} to obtain the explicit computations of $B$ and $D$,
\begin{equation}\label{eqn_D_update}
 \tilde{D}_{i,i} = D_{i,i} + (v^i)^\top M^i v^i   
\end{equation}
and, for all $j<i$,
\begin{equation}\label{eqn_LDL_b_j}
(v^i)^\top(\tilde{D}_{j,j}b^j) =  (v^i)^\top M^j v^j,
\end{equation}
where one choice of $b^j$ can be
\[
    b^j = \frac{1}{\Tilde{D}_{j,j}} M^j v^j.
\]
if $\tilde{D}_{j,j}>0$. 

By taking this choice and summarizing the above equations, we obtain a vector-form $LDL^\top$ decomposition shown in Procedure \ref{procedure_vector_form_LDL}.
\floatname{algorithm}{Procedure}
\begin{algorithm}[h]
    \caption{Vector-form $LDL^\top$ decomposition of \eqref{eqn_LDL_decomposition}}\label{procedure_vector_form_LDL}
    \textbf{Input}: $V, D$;
    \vspace*{.1cm}\hrule\vspace*{.1cm}
        $\textbf{init}~M\leftarrow I_m\in\mathbb{R}^{m\times m}~\text{and}~ B\leftarrow0\in\mathbb{R}^{m\times n}$;\\
        \textbf{for} $i=1,\ldots, n$ \textbf{do},
        \begin{enumerate}[label*=\arabic*., ref=\theenumi{}]
            \item $ q \leftarrow Mv^i$;
            \item $\Tilde{D}_{i,i}\leftarrow D_{i,i} + (v^i)^\top q$;
            \item $ b^i \leftarrow \frac{1}{\Tilde{D}_{i,i}} q$;
            \item $ M\leftarrow M- \Tilde{D}_{i,i} b^i(b^i)^\top$;
        \end{enumerate}
    \textbf{end}.
    \vspace*{.1cm}\hrule\vspace*{.1cm}
    \textbf{Output}: $\Tilde{D}, B$.
\end{algorithm}
\begin{lemma}\label{lemma_M_SPD}
    Given a symmetric positive semidefinite $M\in\mathbb{R}^{m\times m}$ and $\sigma>0$, then, for an arbitrary $v\in\mathbb{R}^m$, the matrix $M^+\triangleq M - \frac{1}{\sigma+v^\top M v}Mvv^\top M^\top$ is symmetric positive semi-definite.
\end{lemma}
\begin{proof}
    Define $\xi\triangleq \sigma + v^\top M v$, which is clearly positive. For an  an arbitrary $x\in\mathbb{R}^m$, we have that
    \[
    \begin{aligned}
        x^\top M^+ x &= x^\top M x - \frac{(x^\top M v)^2}{\sigma+v^\top M v}\\
        &=\frac{\sigma+v^\top M v}{\xi} x^\top M x  - \frac{(x^\top M v)^2}{\xi}\\
        &=\frac{\sigma}{\xi}x^\top M x  + \frac{1}{\xi}\!\left( (x^\top M x)(v^\top M v) - (x^\top M v)^2\right)
    \end{aligned}
    \]
    According to the Cauchy-Schwart inequality and $\xi>0$, we have $(x^\top M x)(v^\top M v) - (x^\top M v)^2\geq0$. Thus, we have
    \[
     x^\top M^+ x \geq\frac{\sigma}{\xi}x^\top M x \geq 0,
    \]
    which completes the proof.
\end{proof}
\begin{corollary}\label{corollary_positive}
        Suppose that $D$ is a positive diagonal matrix, the matrices $M^i$ in Procedure \ref{procedure_vector_form_LDL} are all symmetric positive semi-definite and $\Tilde{D}$ in Procedure \ref{procedure_vector_form_LDL} is a positive diagonal matrix.
\end{corollary}
\begin{proof}
    First, it holds for $M^1$ as $M^1=I_m$, so $\Tilde{D}_{1,1}=D_{1,1}+(v^1)^\top M^1 v^1>0$. Next, by Lemma \ref{lemma_M_SPD}, $M^2$ is symmetric positive semidefinite and then $\Tilde{D}_{2,2}=D_{2,2}+(v^2)^\top M^2 v^2>0$. Recursively repeating this procedure completes the proof.
\end{proof}

\begin{theorem}
The output $\Tilde{D}$ and $B$ of Procedure \ref{procedure_vector_form_LDL} makes the $LDL^\top$ decomposition in \eqref{eqn_LDL_decomposition} unique.
\end{theorem}
\begin{proof}
    By Corollay \ref{corollary_positive}, the choice of $\tilde{D}, B$ in Procedure \ref{procedure_vector_form_LDL} can make \eqref{eqn_v_i_v_i} and \eqref{eqn_v_i_v_j} hold. That is, Procedure \ref{procedure_vector_form_LDL} results in an $LDL^\top$ decomposition, with all elements equal to $VV^\top+D$. Although the choice of $B$ is not unique, the entry of $L(V, B)$ is unique since it is well known that an $LDL^\top$ decomposition of a symmetric positive definite matrix exists and is unique. This completes the proof.
\end{proof}

\subsection{Substitution procedure of parallel vector-form $LDL^\top$}
Like the decomposition procedure, the substitution procedure of our vector-form $LDL^\top$ decomposition is a parallel and vectorized extension of \eqref{eqn_spec_L_solving}, performing all operations at once, unlike \eqref{eqn_spec_L_solving} performing $m$ times.

After performing Procedure \ref{procedure_vector_form_LDL}, the Newton system \eqref{eqn_VV_D_z_u} becomes $L(V,B)\tilde{D}L(V,B)^\top \Delta z = p$, which is equivalent to 
\[
\begin{aligned}
    &\Delta \tilde{z}: L(V,B)\Delta \Tilde{z} =  p\\
    &\Delta \hat{z}: \tilde{D}\Delta \hat{z} = \Delta \tilde{z}\\
    &\Delta z: L(V,B)^\top \Delta z = \Delta \hat{z}
\end{aligned}
\]
Take solving $\Delta \tilde{z}: L(V,B)\Delta \Tilde{z} =  p$ as an example, its solution is 
\[
\begin{aligned}
    \Delta \tilde{z}_1 &= p_1\\
    \Delta\tilde{z}_2 &= p_2 - (v^2)^\top b^1 \Delta \tilde{z}_1\\
    \Delta\tilde{z}_3 &= p_3 - (v^3)^\top b^1 \Delta \tilde{z}_1 - (v^3)^\top b^2 \Delta \tilde{z}_2\\
    &~\vdots\\
    \Delta\tilde{z}_n &= p_n - (v^n)^\top \!\left(\sum_{k=1}^{n-1}b^k\Delta\Tilde{z}_k \right).
\end{aligned}
\]
By defining $q^i\triangleq\sum_{k=1}^{i}b^k\Tilde{z}_k$, whose recurrence is $q^0=0_m, q^{i}=q^{i-1} + \Delta\Tilde{z}_ib^i$, then
\[
\Delta\Tilde{z}_i = p_i - (v^i)^\top q^i,
\]
which clearly can be regarded as a vectorized extension of \eqref{eqn_spec_L_solving}. $\tilde{D}\Delta \hat{z} = \Delta \tilde{z}$ is trivial, $L(V,B)^\top \Delta z = \Delta \hat{z}$ follows the same derivation like $L(V,B)\Delta \Tilde{z} =  p$, and now we summarize their operations in Procedure \ref{procedure_LDL_substitution}.
\floatname{algorithm}{Procedure}
 \begin{algorithm}[h]
    \caption{Solving \eqref{eqn_VV_D_z_u} by using vector-form $LDL^\top$ decomposition}\label{procedure_LDL_substitution}
    \textbf{Input}: $V, \Tilde{D}, B$ and $u$;
    \vspace*{.1cm}\hrule\vspace*{.1cm}
        $\textbf{init}~q \leftarrow 0$;\\
        \textbf{for} $i=1,\ldots, n$ \textbf{do},
        \begin{enumerate}[label*=\arabic*., ref=\theenumi{}]
            \item $\Delta z_i\leftarrow u_i-(v^i)^\top q$;
            \item $q\leftarrow q +\Delta z_i b^i $;
        \end{enumerate}
    \textbf{end}.\\
    \textbf{for} $i=1,\ldots, n$ \textbf{do},
        \begin{enumerate}[label*=\arabic*., ref=\theenumi{}]
        \item [3.] $\Delta z_i\leftarrow \Delta z_i/\tilde{D}_{i,i}$;
        \end{enumerate}
    \textbf{end}.\\
    $\textbf{init}~q\leftarrow 0$;\\
     \textbf{for} $i=n,\ldots, 1$ \textbf{do},
    \begin{enumerate}[label*=\arabic*., ref=\theenumi{}]
        \item [4.] $\Delta z_i\leftarrow\Delta z_i- (b^i)^\top q$;
        \item[5.] $ q \leftarrow q +\Delta z_iv^i$
    \end{enumerate}
    \textbf{end}.
    \vspace*{.1cm}\hrule\vspace*{.1cm}
    \textbf{Output}: $\Delta z$.
\end{algorithm}

Procedure \ref{procedure_vector_form_LDL} and \ref{procedure_LDL_substitution} form a $LDL^\top$ decomposition method in solving \eqref{eqn_VV_D_z_u}, thus inheriting the numerical stability in IPM iterations like the standard Cholesky and product-form $LDL^\top$ decomposition methods \cite{wright1997stability}.

\subsection{Parallel implementation}
The decomposition Procedure \ref{procedure_vector_form_LDL} consists of $n$ iterations, each of which involves matrix-vector multiplication ($O(m^2)$ [flops]), vector-vector inner product, vector assignment, and matrix's rank-1 update  ($O(m^2)$ [flops]), making the computation complexity of Procedure \ref{procedure_vector_form_LDL} $O(nm^2)$ and thus being able to exploit the low-rank feature ($m\leq n$). Moreover, all these operations are \textit{vector-wise}, rather than \textit{element-wise} as in the product-form $LDL^\top$ decomposition (see Appendix \ref{sec_appendix_prodLDL}). Similarly, the substitution Procedure \ref{procedure_LDL_substitution} also consists of $n$ iterations, each of which also involves \textit{vector-wise} vector-vector inner product and vector assignment, making the computation complexity of Procedure  \ref{procedure_LDL_substitution} $O(nm)$.

To exploit the \textit{vector-wise} feature, the implementations of Procedure  \ref{procedure_vector_form_LDL} and \ref{procedure_LDL_substitution} use Advanced Vector Extensions (AVX) instructions (using instruction-level parallelism) to accelerate all vector-vector inner product operations, and loop unrolling or Open Multi-Processing (OpenMP) techniques to parallelize operations on each row of matrices. Therefore, we name our proposed Procedure \ref{procedure_vector_form_LDL} and \ref{procedure_LDL_substitution} as parallel vector-form $LDL^\top$ decomposition. This explains why our parallel vector-form $LDL^\top$ decomposition can be faster than the product-form $LDL^\top$ decomposition, even though it requires more [flops] than the product-form $LDL^\top$ decomposition (noted in Appendix \ref{sec_appendix_block}). Appendix \ref{sec_appendix_block} also points out that our vector-form $LDL^\top$ decomposition can be faster if we use the Blocking technique\footnote{Blocking techniques can accelerate the decomposition of large matrices. All the decomposition methods compared in this paper (i.e., Cholesky decomposition, product-form $LDL^\top$ decomposition, and vector-form $LDL^\top$ decomposition) are non-block versions for a fair comparison.}, which reduces the required [flops] while maintaining vector-wise operations.

Algorithm \ref{alg_time_certifed_IPM}, which integrates our proposed parallel vector-form $LDL^\top$ decomposition into \eqref{procedure_ETC_BoxQP}, is the whole description of our proposed execution-time-certified $\ell_1$-penalty soft-constrained MPC algorithm for closed-loop control

\floatname{algorithm}{Algorithm}
\begin{algorithm}
    \caption{An execution-time certified algorithm for QP (\ref{problem_QP_to_penaltyQP})
    }\label{alg_time_certifed_IPM}
    \textbf{Input}: the problem data $\{Q,F\bar{x}(t),G,g+S\bar{x}(t)\}$, then we can compute $L_Q\leftarrow \mathrm{chol}_L(Q)$, $V\leftarrow GL_Q^{-\top}$ offline;  the number of inequality constraints $n$, then $\eta\leftarrow\frac{\sqrt{2}-1}{\sqrt{2n}+\sqrt{2}-1}$, $\tau\leftarrow\frac{1}{1-\eta}$, and $\lambda\leftarrow\frac{1}{\sqrt{n+1}}$; the stopping tolerance $\epsilon$ (e.g., $1$$\times$$10^{-12}$). Then, we calculate the required exact number of iterations $\mathcal{N}=\left\lceil\frac{\log(\frac{2n}{\epsilon})}{-2\log(\frac{\sqrt{2n}}{\sqrt{2n}+\sqrt{2}-1})}\right\rceil\!+1$.
    \vspace*{.1cm}\hrule\vspace*{.1cm}
    \begin{enumerate}[label*=\arabic*., ref=\theenumi{}]
        \item $d\leftarrow VL_Q^{-1}F\bar{x}(t)+g+S\bar{x}(t),~h\leftarrow \diag(\rho)(VV^\top\rho + 2d)$;
        \item \textbf{if }$\|h\|_\infty=0$, $z\leftarrow0$,
        \textbf{goto} Step 6, \textbf{otherwise},
        \item $V\leftarrow\sqrt{\frac{2\lambda}{\|h\|_\infty}}\diag(\rho)V$;
        \item $(z,\gamma,\theta,\phi,\psi)$ are initialized from (\ref{eqn_initialization_stragegy});
        \item \textbf{for} $k=1, 2,\cdots{}, \mathcal{N}$ \textbf{do}
        \begin{enumerate}[label=\theenumi{}.\arabic*., ref=\theenumi{}.\arabic*]
        \item $\tau\leftarrow(1-\eta)\tau$;
        \item $D \leftarrow \diag\left(\frac{\gamma}{\phi}\right) + \diag\left(\frac{\theta}{\psi}\right)$;
        \item $p \leftarrow 2\left(\sqrt{\frac{\theta}{\psi}}\tau e-\sqrt{\frac{\gamma}{\phi}}\tau e+ \gamma - \theta\right)$;
        \item calculate $\Delta z$ via Procedure \ref{procedure_vector_form_LDL} and Procedure \ref{procedure_LDL_substitution};
        \item calculate $(\Delta\gamma,\Delta\theta,\Delta\phi,\Delta\psi)$ from \eqref{eqn_Delta_gamma_theta_phi_psi};
        \item  $z\leftarrow z+\Delta z$, $\gamma\leftarrow \gamma+\Delta \gamma$, $\theta\leftarrow \theta+\Delta \theta$, $\alpha\leftarrow \phi+\Delta \phi$, $\omega\leftarrow \psi+\Delta \psi$;
        \end{enumerate}
        \item[~] \textbf{end}
        \item  return $y^*=-\left(L_QL_Q^\top \right)^{-1}\!\left(F\bar{x}(t)+\frac{1}{2}G^\top(\rho z+\rho)\right)$.
    \end{enumerate}
\end{algorithm}

\section{Numerical Experiments}
The reported numerical results were obtained on a Dell Precision 7760 (equipped with 2.5~GHz 16-core Intel Core i7-11850H and 128GB RAM) running Julia v1.10 on Ubuntu 22.04. Algorithm~\ref{alg_time_certifed_IPM} is implemented in C code and executed in Julia via Julia-C interface \textit{ccall}. Code is available at \url{https://github.com/liangwu2019/L1-penalty-QP}.

\subsection{Speedup evaluation of parallel vector-form $LDL^\top$ decomposition}
\begin{figure*}[!t]
    \centering
    \begin{minipage}[b]{0.32\linewidth}
        \captionsetup{labelformat=empty, justification=centering, font=small}
        \includegraphics[width=\linewidth]{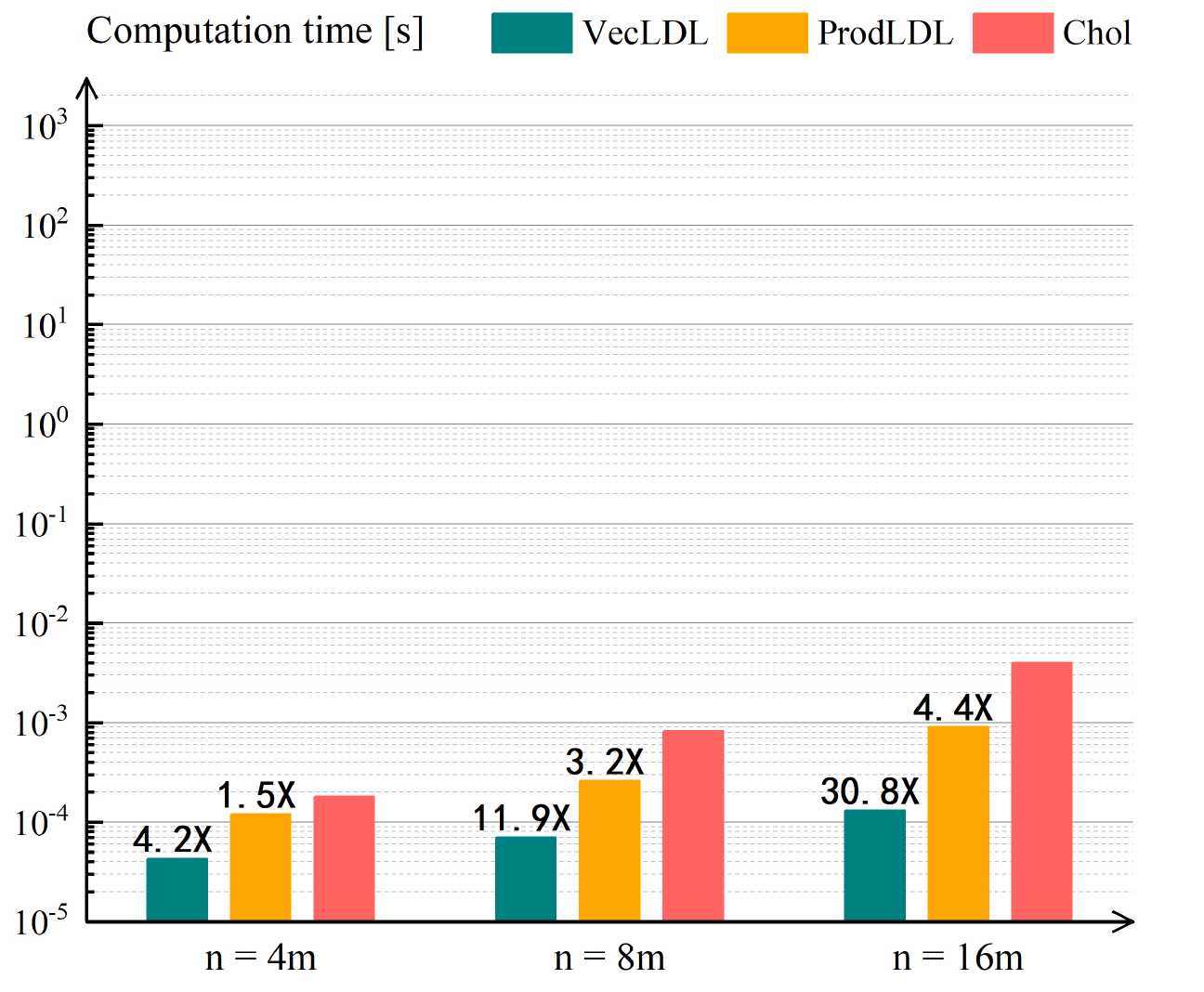}
        \caption*{(a) $m=25$}
    \end{minipage}
    \hfill
    \begin{minipage}[b]{0.32\linewidth}
        \captionsetup{labelformat=empty, justification=centering, font=small}
        \includegraphics[width=\linewidth]{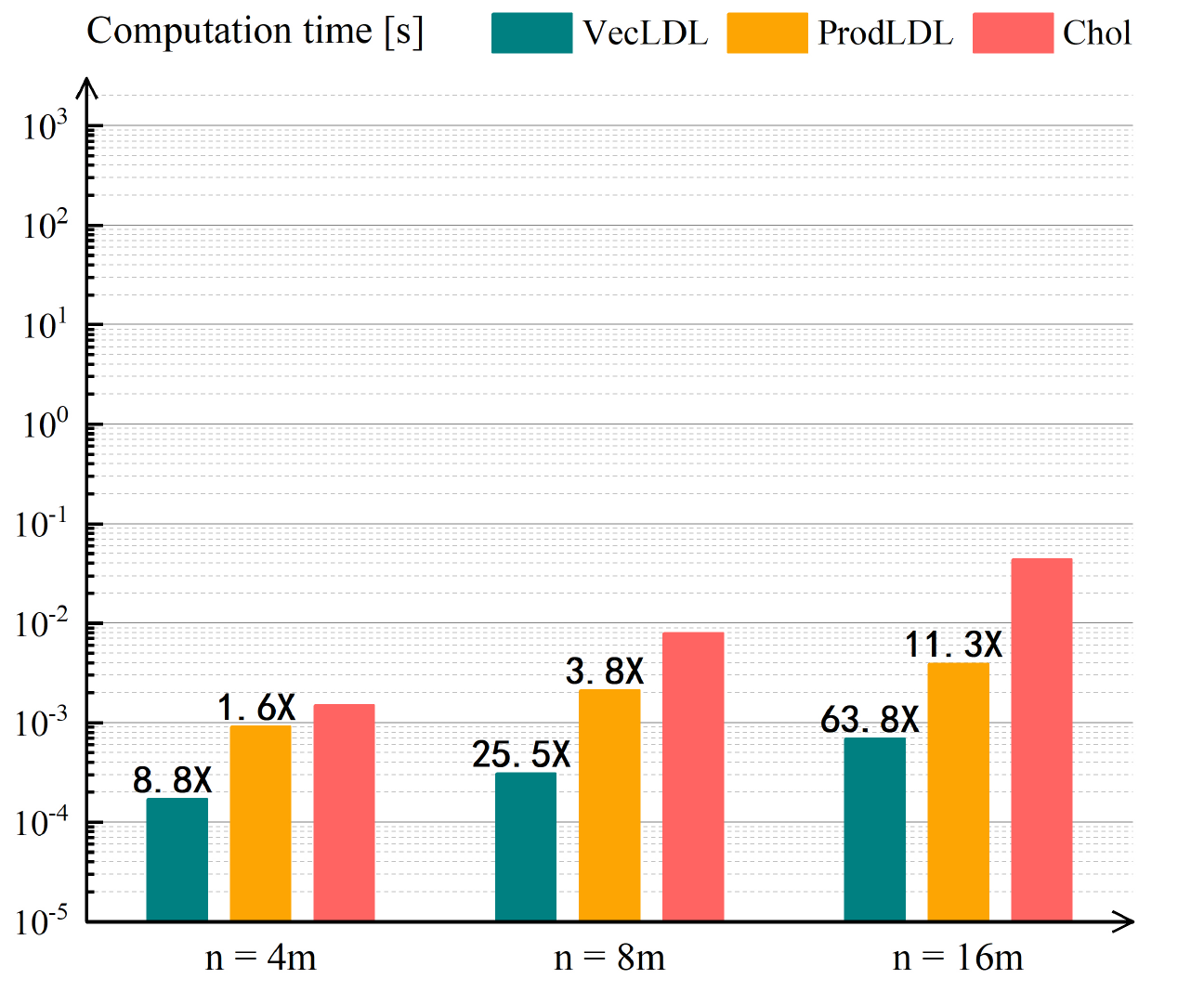}
        \caption*{(b) $m=50$}
    \end{minipage}
    \hfill
    \begin{minipage}[b]{0.32\linewidth}
        \captionsetup{labelformat=empty, justification=centering, font=small}
        \includegraphics[width=\linewidth]{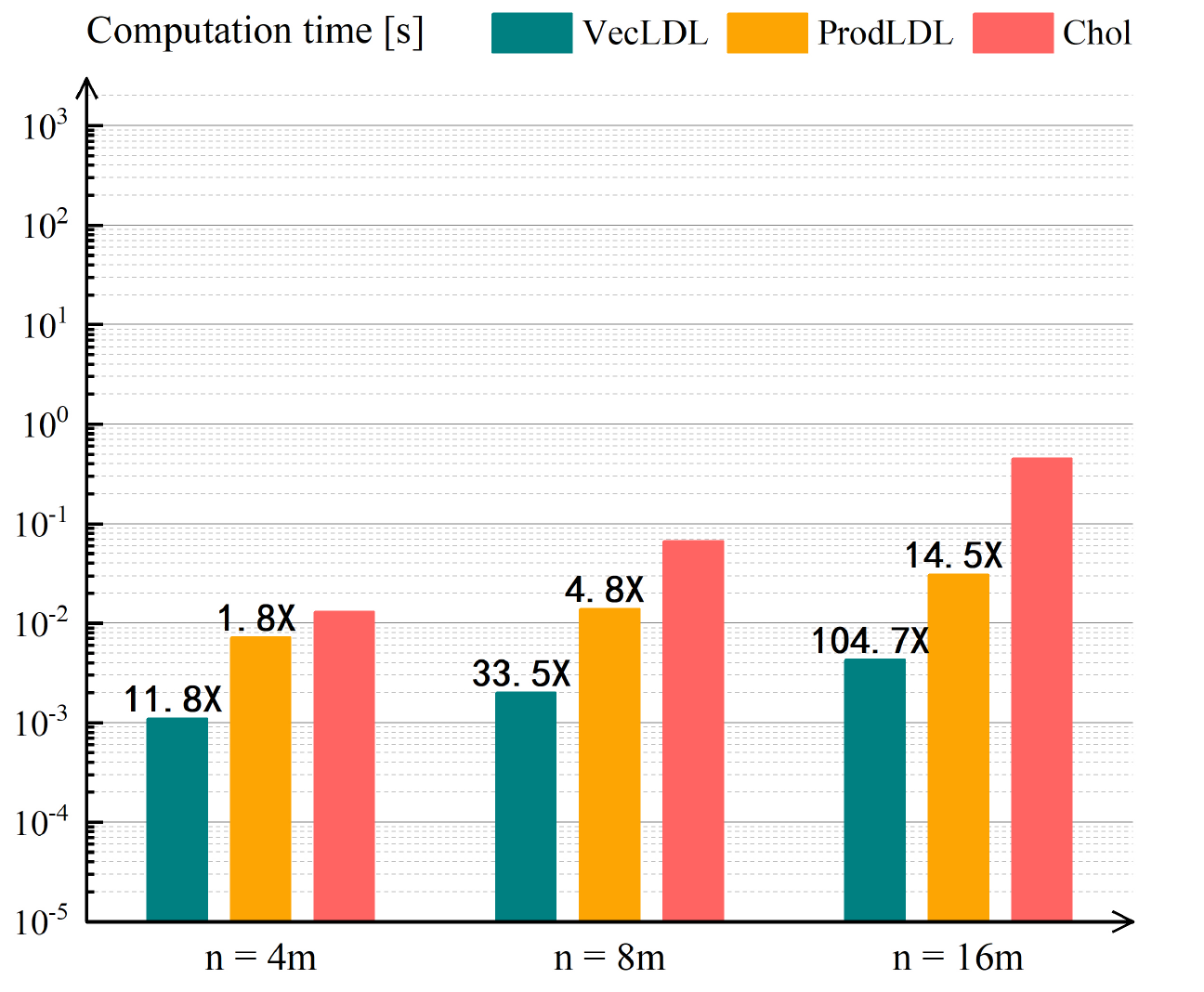}
        \caption*{(c) $m=100$}
    \end{minipage}
    
    \vspace{5mm}
    
    \begin{minipage}[b]{0.32\linewidth}
        \captionsetup{labelformat=empty, justification=centering, font=small}
        \includegraphics[width=\linewidth]{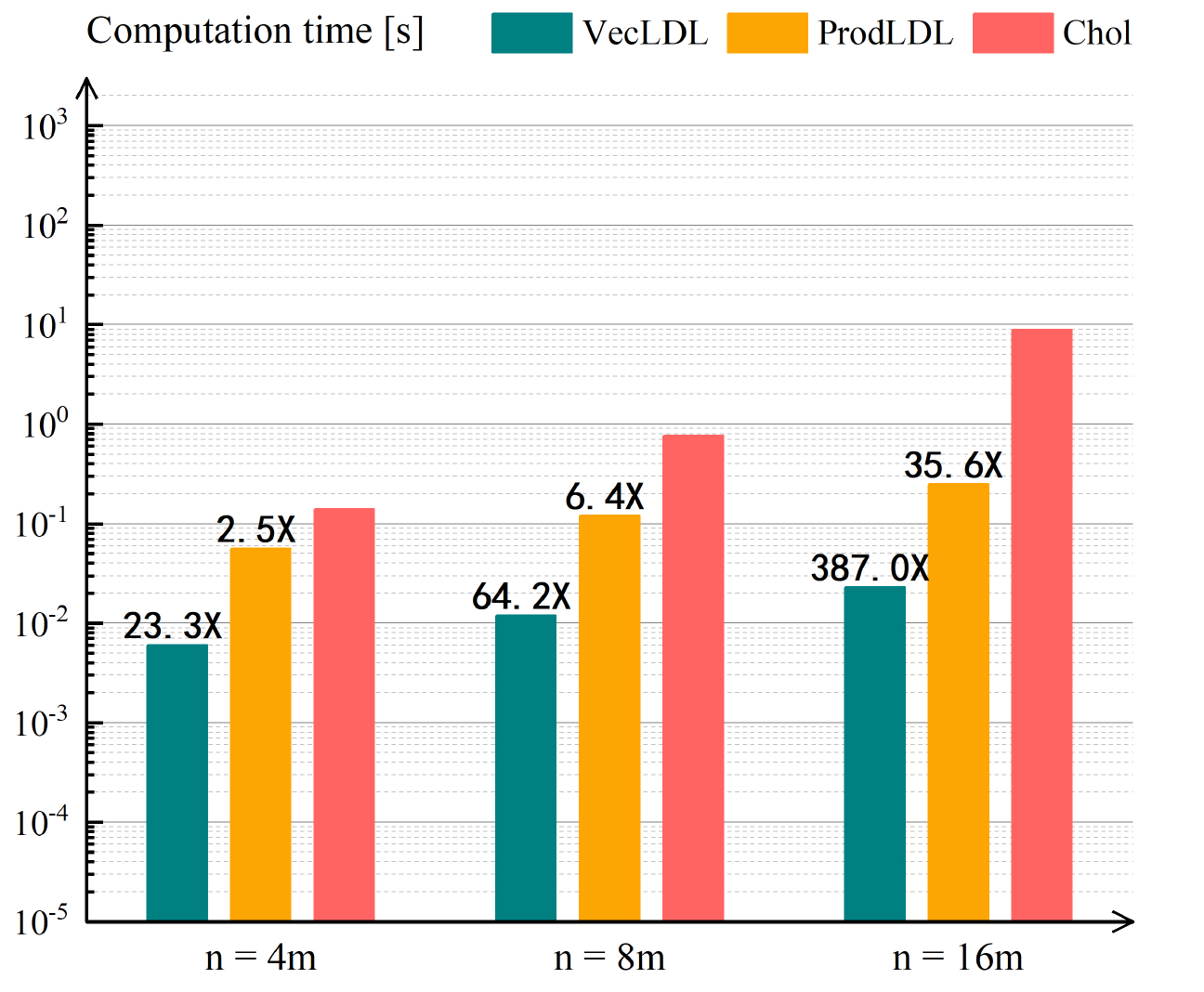}
        \caption*{(d) $m=200$}
    \end{minipage}
    \hfill
    \begin{minipage}[b]{0.32\linewidth}
        \captionsetup{labelformat=empty, justification=centering, font=small}
        \includegraphics[width=\linewidth]{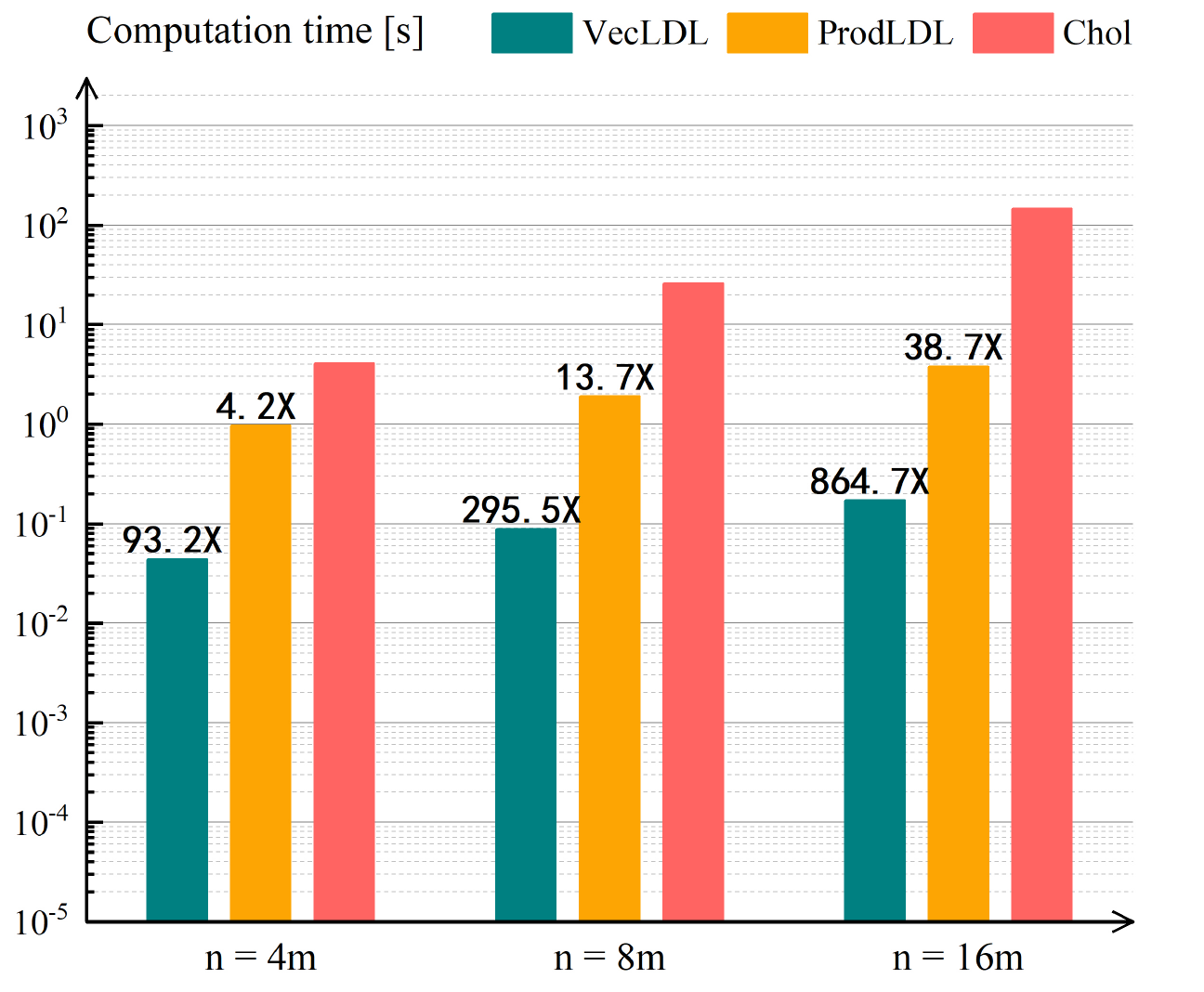}
       \caption*{(e) $m=500$}
    \end{minipage}
    \hfill
    \begin{minipage}[b]{0.32\linewidth}
        \captionsetup{labelformat=empty, justification=centering, font=small}
        \includegraphics[width=\linewidth]{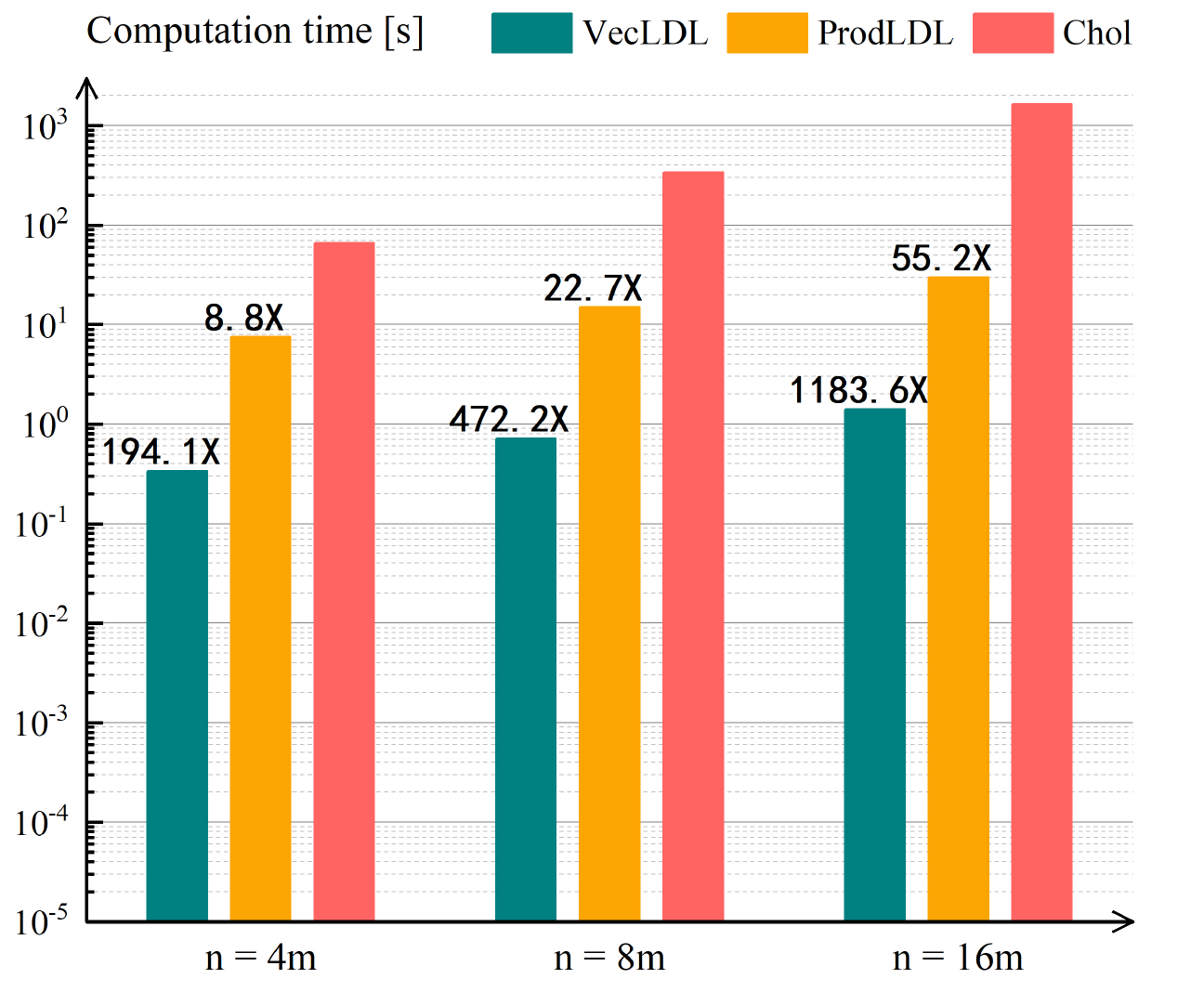}
        \caption*{(f) $m=1000$}
    \end{minipage}
    \caption{The computation time of VecLDL, ProdLDL, and Chol implementations in solving $(VV^\top+D)\Delta z = p$ on different problem sizes $(m,n)$.}\label{fig1}
\end{figure*}

To validate the computation advantage in solving $(VV^\top+D)\Delta z = p$, we compare the computation time of the parallel vector-form $LDL^\top$ decomposition developed in the previous section (labeled as VecLDL) against the standard Cholesky decomposition and the Product-form $LDL^\top$ decomposition methods (labeled as Chol and ProdLDL, respectively), which are also implemented in C code and executed in Julia. Fig. \ref{fig1} plots the CPU time of VecLDL, ProdLDL, and Chol methods on random linear systems $(VV^\top+D)\Delta z = p$ with different sizes $(m,n)$ such as $n=4m$, $n=8m$, $n=16m$, and $m$ varying from 25 to 1000. 
The speedup factors of VecLDL and ProdLDL in Fig. \ref{fig1} are calculated by dividing their computation time by the Chol computation time. Fig. \ref{fig1} shows that VecLDL is the most computationally efficient, the speedup factor gradually increases with the increase of $\frac{n}{m}$ and $m$, even up to 1000-fold for $m=1000,n=16m$.

\subsection{Accuracy of $\ell_1$-penalty soft-constrained QP}
To validate the solution accuracy of the $\ell_1$-penalty soft-constrained QP scheme \eqref{problem_QP_to_penaltyQP}, we consider QPs with different condition numbers of $Q$ (from $10^1$ to $10^6$) and feasible solution (by setting $m=20,n=8m$ and the constraint matrix $G=\mathrm{sprandn(n,m,0.8)}$). We choose $\rho$ as same as the condition number of $Q$. For each scenario with a specified condition number of $Q$, we randomly generate 50 QPs and apply Algorithm \ref{alg_time_certifed_IPM}, OSQP.jl (wrapper of OSQP solver \cite{stellato2020osqp}), and Ipopt.jl (wrapper of IPOPT solver \cite{wachter2006implementation}).

Except that OSQP suffers in achieving high-accuracy solutions even though here we turn on its solver settings as $\mathrm{polishing}=1,\mathrm{eps\_abs}=\mathrm{eps\_rel}=10^{-20}, \mathrm{max\_iter}=10^5$. Thus, Fig. \ref{fig2}a does not plot the results of OSQP. Algorithm \ref{alg_time_certifed_IPM} and IPOPT can both achieve nearly $10^{-6}$ worst-case solution accuracy on different condition number instances varying from $10$ to $10^6$, by setting the convergence tolerance of Algorithm \ref{alg_time_certifed_IPM} and IPOPT as $10^{-20}$ and $10^{-12}$, respectively, as shown in Fig. \ref{fig2}a,

Fig. \ref{fig2}a also shows that Algorithm \ref{alg_time_certifed_IPM} has a smaller worst-case computation time than IPOPT. Moreover, the computation time of Algorithm \ref{alg_time_certifed_IPM} does not vary with the condition number of $Q$, unlike IPOPT. This is consistent with our theoretical results, namely that Algorithm \ref{alg_time_certifed_IPM} have \textit{data-independent} computation complexity.

\subsection{AFTI-F16 MPC problem with initial infeasibility}
In order to test Algorithm \ref{alg_time_certifed_IPM} in an MPC problem,  we consider the open-loop unstable AFTI-F16 aircraft example described in \cite{bemporad1997nonlinear},
\[
\left\{\begin{aligned}
\dot{x} =&{\footnotesize\left[\begin{array}{cccc}
-0.0151 & -60.5651 & 0 & -32.174 \\
-0.0001 & -1.3411 & 0.9929 & 0 \\
0.00018 & 43.2541 & -0.86939 & 0 \\
0 & 0 & 1 & 0
\end{array}\right]} x\\&+{\footnotesize\left[\begin{array}{cc}
-2.516 & -13.136 \\
-0.1689 & -0.2514 \\
-17.251 & -1.5766 \\
0 & 0
\end{array}\right] }u \\
y =&{\footnotesize\left[\begin{array}{llll}
0 & 1 & 0 & 0 \\
0 & 0 & 0 & 1
\end{array}\right]x}
\end{aligned}\right.
\]
which is sampled using zero-order hold every 0.05~s. The control goal is to regulate the attach angle $y_1$ to zero and make the pitch angle $y_2$ track a reference signal $r_2$ under the input constraints $|u_i| \leq 25^{\circ}$, $i = 1, 2$ and the output constraints $-0.5\leq y_1 \leq 0.5 $ and $-100 \leq y_2 \leq 100$. 

If the system suffers a sudden disturbance making the current state violate the output constraint, then obviously the resulting QP  has no feasible solution. For example, when $x(0)=[0,5,0,0]^\top$ ($y_1(0)=x_2(0)=5$), the constraint $-0.5\leq y_1 \leq 0.5$ is violated. To circumvent this infeasibility, we adopt the $\ell_1$-penalty soft-constrained QP scheme \eqref{problem_QP_to_penaltyQP} with $\rho=10^3$ or $10^4$ for output and input constraints, and the MPC cost weight $W_y = \diag$([10,10]), $W_u = 0$, $W_{\Delta u}= \diag$([0.1, 0.1]), and the prediction horizon is $Np=10$. The size of resulting $\ell_1$-penalty soft-constrained QP is $m=20,n=4m$. 

\begin{figure*}[!t]
    \centering
    \begin{minipage}[b]{0.54\linewidth}
        \captionsetup{labelformat=empty, justification=centering, font=small}
        \includegraphics[width=\linewidth]{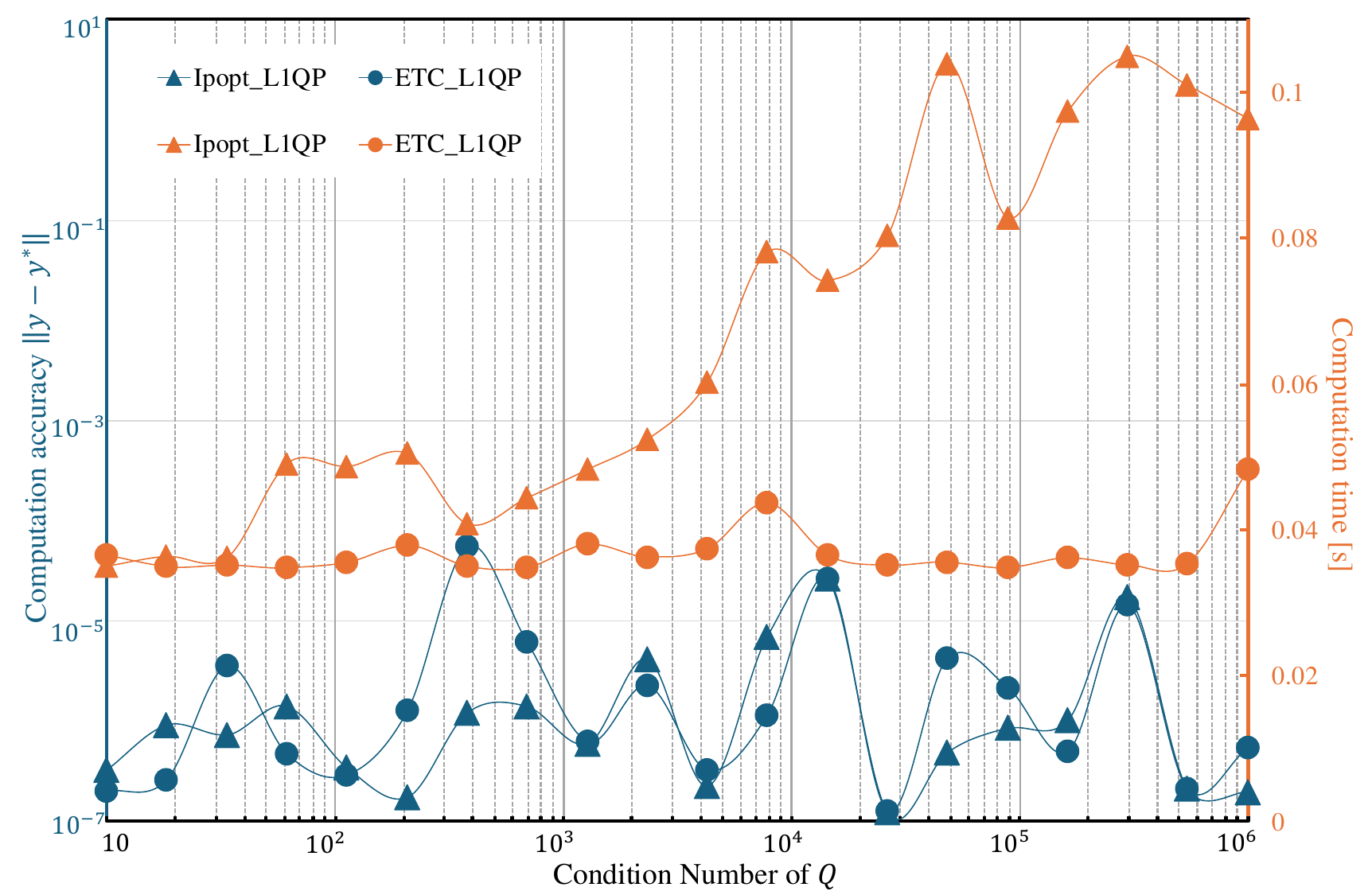}
        \caption*{(a)}
    \end{minipage}
    \hfill
    \begin{minipage}[b]{0.45\linewidth}
        \captionsetup{labelformat=empty, justification=centering, font=small}
        \includegraphics[width=\linewidth]{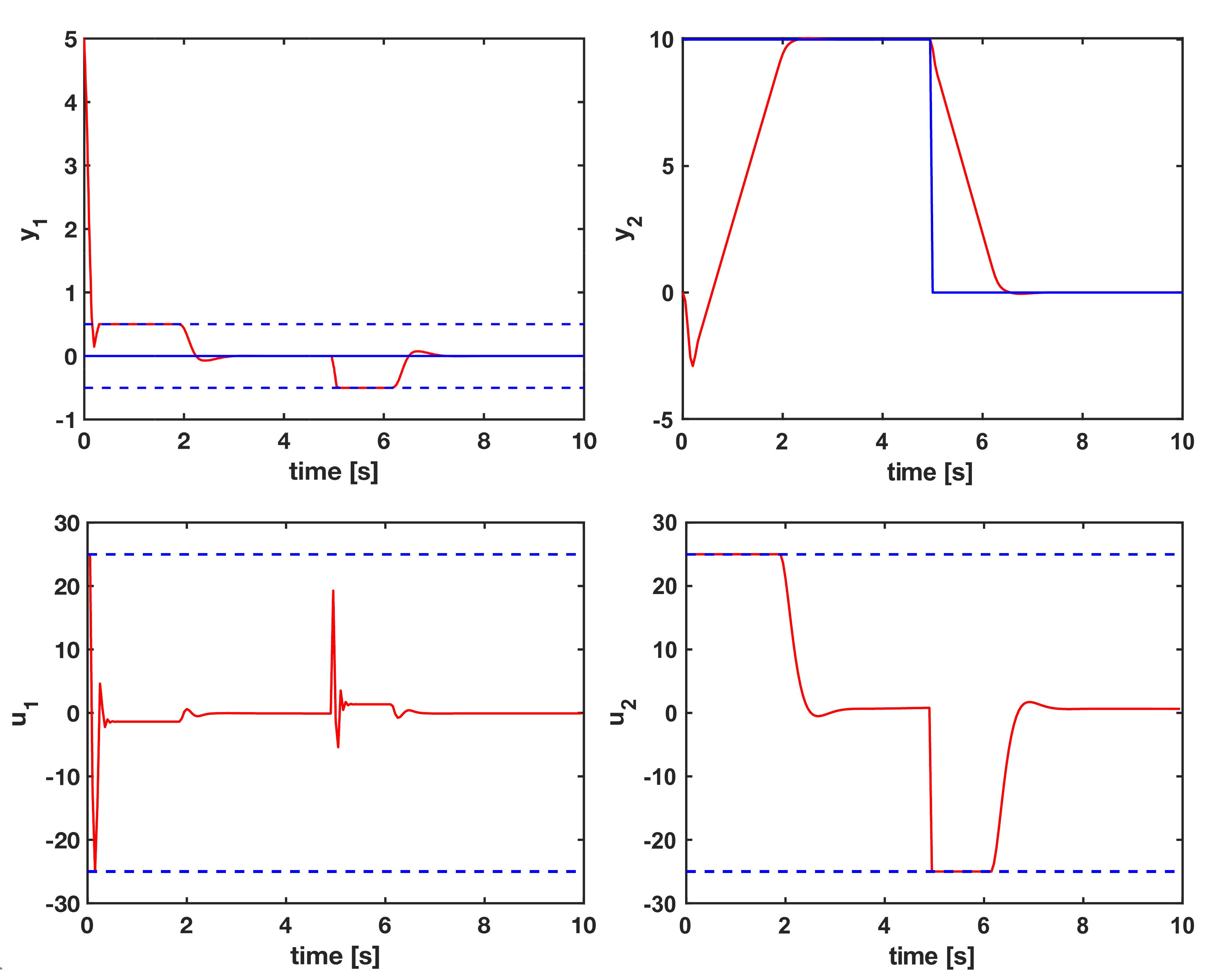}
        \caption*{(b)}
    \end{minipage}
    \caption{Results of numerical examples. (a) Comparison (worst-case computation accuracy $\|y-y^*\|$ and worst-case computation time) of  Algorithm \ref{alg_time_certifed_IPM} and IPOPT and over random QPs ($m=20,n=8m$, transformed into $\ell_1$-penalty QP) as a function of the condition number of $Q$. (b) AFTI-F16 closed-loop performance by adopting $\ell_1$-penalty soft-constrained MPC (the values of $\rho$ for output and input constraints are $10^3$ and $10^4$, respectively) using Algorithm \ref{alg_time_certifed_IPM}.}\label{fig2}
\end{figure*}

The AFTI-F16 closed-loop performance is shown in Fig. \ref{fig2}b, which tells that quickly pulling the attack angle $y_1$ back to the safe region is completed as a first priority. 
The pitch angle $y_2$ first moves in the opposite direction of the desired tracking signal and only starts to move towards the desired tracking signal when $y_1$ enters the safe region. The physical input constraints are never violated thanks to the larger value of $\rho$ for input constraints and the sparseness of $\ell_1$-penalty soft-constrained scheme.

Furthermore, we compare the worst-case computation time of Algorithm \ref{alg_time_certifed_IPM}, IPOPT, and OSQP. As our considered AFTI-F16 MPC problem has initial infeasibility, IPOPT and OSQP are also used to solve globally feasible $\ell_1$-penalty soft-constrained QP \eqref{problem_QP_to_penaltyQP} (namely Box-QP \eqref{problem_box_QP_0}). Under the premise of generating the same closed-loop performance, we choose the default settings of IPOPT and OSQP solvers and we set the tolerance $\epsilon$ of Algorithm \ref{alg_time_certifed_IPM} as $10^{-6}$. The worst-case computation times of the three solvers among different prediction horizon length scenarios are listed in Table \ref{tab:worst_case_time}, which shows that Algorithm \ref{alg_time_certifed_IPM} has the smallest worst-case computation time 
in solving the $\ell_1$-penalty soft-constrained MPC problems.

\begin{table}[!htbp]
    \caption{Worst-case computation time [s] of different solvers}
    \centering
    \begin{tabular}{clll}
    \toprule
     Prediction horizon length  & Algorithm \ref{alg_time_certifed_IPM} & IPOPT-$\ell_1$QP & OSQP-$\ell_1$QP\\
     \midrule
      $N=5$  & $\mathbf{0.0018}$ & 0.0082 & 0.0075 \\
      $N=10$ & $\mathbf{0.0055}$ & 0.0173 & 0.0249\\
      $N=15$ & $\mathbf{0.0232}$ & 0.0302 & 0.0522\\
      $N=20$ & $\mathbf{0.0342}$ & 0.0394  & 0.1086\\
     \bottomrule
    \end{tabular}
    \label{tab:worst_case_time}
\end{table}

\section{Conclusion}
This paper proposes an $\ell_1$-penalty soft-constrained MPC formulation to handle possible infeasibility and provide an execution time certificate for real-time closed-loop MPC. The $\ell_1$-penalty soft-constrained MPC can be surprisingly transformed into a Box-QP and then our previous Box-QP algorithm \cite{wu2023direct} can be applied to provide a theoretical execution-time certificate. Moreover, this paper proposes a parallel vector-form $LDL^\top$ decomposition for the first time, to accelerate the most computationally expensive Newton step, which enables our solver to achieve comparable computation performance to the state-of-the-art solvers such as IPOPT and OSQP. Future work includes implementing a block-version of our parallel vector-form $LDL^\top$ decomposition to further improve computation efficiency.

In addition to MPC applications, our globally feasible and execution-time-certified QP Algorithm \ref{alg_time_certifed_IPM} can be applied to other various real-time QP applications, such as constrained Differential Dynamic Programming (DDP) \cite{xie2017differential}, constrained Control Lyapunov policy \cite{wang2010fast}, and Control Barrier Function based QP \cite{ames2016control,xiao2021high}. Moreover, our parallel vector-form $LDL^\top$ decomposition can also benefit other various numerical computation tasks.

\section{Acknowledgement}
This research was supported by the U.S. Food and Drug Administration under the FDA BAA-22-00123 program, Award Number 75F40122C00200.

\appendix
\subsection{Computing procedure of product-form $LDL^\top$ decomposition}\label{sec_appendix_prodLDL}
The derivation is based on the recursive rank-1 update of $LDL^\top$, let assume that we have obtained 
\[
\sum_{k=1}^{l-1}v^k(v^k)^\top +D = L^1\cdots L^{l-1}D^{l-1}(L^{l-1})^\top\cdots(L^1)^\top
\]
then we have
\[
\begin{aligned}
&\quad L^1\cdots L^{l-1}D^{l-1}(L^{l-1})^\top\cdots(L^1)^\top + v^l(v^l)^\top \\
&= L^1\cdots L^{l-1}(D^{l-1}+\tilde{v}^l(\tilde{v}^l)^\top)(L^{l-1})^\top\cdots(L^1)^\top \\
&= L^1\cdots L^{l}D^{l}(L^{l})^\top\cdots(L^1)^\top,
\end{aligned}
\]
where $\tilde{v}^l\in\mathbb{R}^n$ is the solution of $L^1\cdots L^{l-1}\tilde{v}^{l} = v^l$ and $D^{l-1}+\tilde{v}^l(\tilde{v}^l)^\top=L^lD^{l}(L^l)^\top$.

Let's write the element-wise notation of $D^{l-1}+\tilde{v}^l(\tilde{v}^l)^\top$ as
\begin{subequations}
    \begin{eqnarray}
        \tilde{v}^l_i \tilde{v}^l_i + D^{l-1}_{i,i} = D^l_{i,i}+ \sum_{k=1}^{i-1}D^l_{k,k} (\tilde{v}^l_i b^l_k) (\tilde{v}^l_ib^l_k)\\
        \tilde{v}^l_i \tilde{v}^l_j = D^l _{j,j} \tilde{v}^l_i b^l_j+\sum_{k=1}^{j-1}D^l_{k,k} (\tilde{v}^l_i b^l_k) (\tilde{v}^l_jb^l_k),~\text{for}~j<i.
    \end{eqnarray}
\end{subequations}
According to \cite{bennett1965triangular, fletcher1974modification, gill1975methods}, the authors in \cite{fine2001efficient} defined $\kappa_i \triangleq 1-\sum_{k=1}^{i-1}b^l_kb^l_k$ (its recurrence form is $\kappa_1=1, \kappa_{i+1} = \kappa_i -D^l_{i,i}b^l_ib^l_i$ ) and derived the computation of $D^l$ and $b^l_1,\cdots{},b^l_{n-1}$: 
\begin{equation}\label{eqn_prodLDL_update_1}
  \begin{aligned}
  &\kappa_1 = 1\\
  &\textbf{For}~i=1,\cdots{}, n\\
    &\quad D^l_{i,i} = D^{l-1}_{i,i} + \tilde{v}^l_i \kappa_i \tilde{v}^l_i \\
    &\quad b^l_i =\frac{\kappa_i\tilde{v}^l_i}{D^l_{i,i}}\\
    &\quad \kappa_{i+1} = \kappa_i -D^l_{i,i}b^l_ib^l_i
\end{aligned}  
\end{equation}
which only requires $O(n)$ [flops]. In addition, the $l$-th step of the product-form $LDL^\top$ decomposition requires computing $L^1\cdots L^{l-1}\tilde{v}^{l} = v^l$, which is equivalent to sequentially solving $l-1$ triangular systems such as $L^{l-1}\Delta x = \Delta y$. Solving $L^{l-1}\Delta x = \Delta y$ only requires $O(n)$ [flops] by exploiting the special structure of $L^{l-1}$,
\begin{equation}\label{eqn_spec_L_solving}
\begin{aligned}
    &\textbf{Given}~L^{l-1}(\tilde{v}^{l-1},b^{l-1})~\text{and}~ \Delta y:\\
    &\sigma \leftarrow 0\\
    &\textbf{For}~j=1,\cdots{},n\\
    &\quad\quad \Delta x_j\leftarrow \Delta y_j - \tilde{v}^{l-1}_j \sigma\\
    &\quad\quad \sigma \leftarrow \sigma + \Delta x_j b^{l-1}_j
\end{aligned}
\end{equation}
So the decomposition phase and substitution phase of the product-form $LDL^\top$ decomposition totally requires solving $\frac{m(m-1)}{2}$ and $2m$
systems of the form $L^l\Delta x= \Delta y$, respectively.

\subsection{Block version of parallel vector-form $LDL^\top$ and its relationship with product-form $LDL^\top$}\label{sec_appendix_block}
Let's take two blocks as an example.  Given a $V\in\mathbb{R}^{n\times m}$, split as $V=[V^1,V^2]$ where $V^1\in\mathbb{R}^{n\times m_1}$, $V^2\in\mathbb{R}^{n\times m_2}$ and $m_1+m_2=m$. Then
\[
VV^\top +D = V^1(V^1)^\top + V^2(V^2)^\top + D
\]
and we can first apply our proposed parallel vector-form $LDL^\top$ decomposition (namely Procedure \ref{procedure_vector_form_LDL}) for obtaining 
\[
V^1(V^1)^\top+D=L^1D^1(L^1)^\top.
\]
After that, define $\hat{V}^2\triangleq (L^1)^{-1}V^2$, which can be calculated via using Steps 1 and 2 of Procedure \ref{procedure_LDL_substitution} in a parallel way. Apply Procedure \ref{procedure_vector_form_LDL} again for obtaining
\[
\hat{V}^2(\hat{V}^2)^\top + D^1=L^2D^2(L^2)^\top.
\]
Finally, based on calculated $VV^\top +D=L^1L^2D^2(L^2)^\top(L^1)^\top$, apply Procedure \ref{procedure_LDL_substitution} twice to obtain the solution of $(VV^\top+D)\Delta z=p$. 

By summarizing the above analysis and assuming $m_1=m_2=\frac{m}{2}$, we can calculate the exact [flops] required by product-form, vector-form, and 2-block vector-form $LDL^\top$ decompositions, see Table \ref{tab_flops}. It shows that the [flops] of 2-block vector-form $LDL^\top$ decomposition is between product-form and vector-form $LDL^\top$ decompositions while maintaining vector-wise operations. Moreover, the $m$-block vector-form  $LDL^\top$ decomposition is reduced to product-form $LDL^\top$ decomposition at the cost of losing the vector-wise feature and becoming element-wise. Thus, choosing the number of blocks appropriately while maintaining vector-wise operations can achieve a faster implementation. 
\begin{table}[!htbp]
    \caption{[flops] analysis of product-form, vector-form, and 2-block vector-form $LDL^\top$ decompositions}
    \centering
    \begin{tabular}{cc}
    \toprule
      decomposition methods & [flops]\\
     \midrule
      ProdLDL  & $2nm^2+13nm+4n$  \\
      VecLDL & $5nm^2+11nm+4n$ \\
      2-block VecLDL & $\frac{9}{2}nm^2+12nm+7n$\\
     \bottomrule
    \end{tabular}
    \label{tab_flops}
\end{table}

\bibliographystyle{IEEEtran}
\bibliography{ref} 

\begin{thebibliography}{10}
\providecommand{\url}[1]{#1}
\csname url@samestyle\endcsname
\providecommand{\newblock}{\relax}
\providecommand{\bibinfo}[2]{#2}
\providecommand{\BIBentrySTDinterwordspacing}{\spaceskip=0pt\relax}
\providecommand{\BIBentryALTinterwordstretchfactor}{4}
\providecommand{\BIBentryALTinterwordspacing}{\spaceskip=\fontdimen2\font plus
\BIBentryALTinterwordstretchfactor\fontdimen3\font minus \fontdimen4\font\relax}
\providecommand{\BIBforeignlanguage}[2]{{%
\expandafter\ifx\csname l@#1\endcsname\relax
\typeout{** WARNING: IEEEtran.bst: No hyphenation pattern has been}%
\typeout{** loaded for the language `#1'. Using the pattern for}%
\typeout{** the default language instead.}%
\else
\language=\csname l@#1\endcsname
\fi
#2}}
\providecommand{\BIBdecl}{\relax}
\BIBdecl

\bibitem{wu2023direct}
L.~Wu and R.~D. Braatz, ``A direct optimization algorithm for input-constrained {MPC},'' \emph{IEEE Transactions on Automatic Control}, 2024, in press,arXiv:2306.15079.

\bibitem{bemporad2002explicit}
A.~Bemporad, M.~Morari, V.~Dua, and E.~N. Pistikopoulos, ``The explicit linear quadratic regulator for constrained systems,'' \emph{Automatica}, vol.~38, no.~1, pp. 3--20, 2002.

\bibitem{richter2011computational}
S.~Richter, C.~N. Jones, and M.~Morari, ``Computational complexity certification for real-time {MPC} with input constraints based on the fast gradient method,'' \emph{IEEE Transactions on Automatic Control}, vol.~57, no.~6, pp. 1391--1403, 2011.

\bibitem{bemporad2012simple}
A.~Bemporad and P.~Patrinos, ``Simple and certifiable quadratic programming algorithms for embedded linear model predictive control,'' \emph{IFAC Proceedings Volumes}, vol.~45, no.~17, pp. 14--20, 2012.

\bibitem{giselsson2012execution}
P.~Giselsson, ``Execution time certification for gradient-based optimization in model predictive control,'' in \emph{51st IEEE Conference on Decision and Control}, 2012, pp. 3165--3170.

\bibitem{cimini2017exact}
G.~Cimini and A.~Bemporad, ``Exact complexity certification of active-set methods for quadratic programming,'' \emph{IEEE Transactions on Automatic Control}, vol.~62, no.~12, pp. 6094--6109, 2017.

\bibitem{cimini2019complexity}
------, ``Complexity and convergence certification of a block principal pivoting method for box-constrained quadratic programs,'' \emph{Automatica}, vol. 100, pp. 29--37, 2019.

\bibitem{arnstrom2019exact}
D.~Arnstr{\"o}m and D.~Axehill, ``Exact complexity certification of a standard primal active-set method for quadratic programming,'' in \emph{58th IEEE Conference on Decision and Control}, 2019, pp. 4317--4324.

\bibitem{arnstrom2020complexity}
D.~Arnstr{\"o}m, A.~Bemporad, and D.~Axehill, ``Complexity certification of proximal-point methods for numerically stable quadratic programming,'' \emph{IEEE Control Systems Letters}, vol.~5, no.~4, pp. 1381--1386, 2020.

\bibitem{arnstrom2021unifying}
D.~Arnstr{\"o}m and D.~Axehill, ``A {U}nifying {C}omplexity {C}ertification {F}ramework for {A}ctive-{S}et {M}ethods for {C}onvex {Q}uadratic {P}rogramming,'' \emph{IEEE Transactions on Automatic Control}, vol.~67, no.~6, pp. 2758--2770, 2021.

\bibitem{okawa2021linear}
I.~Okawa and K.~Nonaka, ``Linear complementarity model predictive control with limited iterations for box-constrained problems,'' \emph{Automatica}, vol. 125, p. 109429, 2021.

\bibitem{wu2024time}
L.~Wu, K.~Ganko, and R.~D. Braatz, ``Time-certified {I}nput-constrained {NMPC} via {K}oopman operator,'' in \emph{8th IFAC Conference on Nonlinear Model Predictive Control}, 2024, in press, arXiv:2401.04653.

\bibitem{wu2024execution}
L.~Wu, K.~Ganko, S.~Wang, and R.~D. Braatz, ``An {E}xecution-time-certified {R}iccati-based {IPM} algorithm for {RTI}-based {I}nput-constrained {NMPC},'' in \emph{63nd IEEE Conference on Decision and Control}, 2024, in press, arXiv:2402.16186.

\bibitem{kerrigan2000soft}
E.~C. Kerrigan and J.~M. Maciejowski, ``Soft constraints and exact penalty functions in model predictive control,'' in \emph{Proceedings of the UKACC International Conference}, Cambridge, UK, September 2000.

\bibitem{wang2009fast}
Y.~Wang and S.~Boyd, ``Fast model predictive control using online optimization,'' \emph{IEEE Transactions on Control Systems Technology}, vol.~18, no.~2, pp. 267--278, 2009.

\bibitem{ferreau2014qpoases}
H.~Ferreau, C.~Kirches, A.~Potschka, H.~Bock, and M.~Diehl, ``qp{OASES}: {A} parametric active-set algorithm for quadratic programming,'' \emph{Mathematical Programming Computation}, vol.~6, pp. 327--363, 2014.

\bibitem{stellato2020osqp}
B.~Stellato, G.~Banjac, P.~Goulart, A.~Bemporad, and S.~Boyd, ``{OSQP}: {A}n operator splitting solver for quadratic programs,'' \emph{Mathematical Programming Computation}, vol.~12, no.~4, pp. 637--672, 2020.

\bibitem{wu2023simple}
L.~Wu and A.~Bemporad, ``A {S}imple and {F}ast {C}oordinate-{D}escent {A}ugmented-{Lagrangian} {S}olver for {M}odel {P}redictive {C}ontrol,'' \emph{IEEE Transactions on Automatic Control}, vol.~68, no.~11, pp. 6860--6866, 2023.

\bibitem{wu2023construction}
------, ``A construction-free coordinate-descent augmented-{L}agrangian method for embedded linear {MPC} based on {ARX} models,'' \emph{IFAC-PapersOnLine}, vol.~56, no.~2, pp. 9423--9428, 2023.

\bibitem{gros2020linear}
S.~Gros, M.~Zanon, R.~Quirynen, A.~Bemporad, and M.~Diehl, ``From linear to nonlinear {MPC}: {B}ridging the gap via the real-time iteration,'' \emph{International Journal of Control}, vol.~93, no.~1, pp. 62--80, 2020.

\bibitem{forgione2020efficient}
M.~Forgione, D.~Piga, and A.~Bemporad, ``Efficient calibration of embedded {MPC},'' \emph{IFAC-PapersOnLine}, vol.~53, no.~2, pp. 5189--5194, 2020.

\bibitem{jerez2011condensed}
J.~L. Jerez, E.~C. Kerrigan, and G.~A. Constantinides, ``A condensed and sparse {QP} formulation for predictive control,'' in \emph{50th IEEE Conference on Decision and Control and European Control Conference}, 2011, pp. 5217--5222.

\bibitem{saraf2017fast}
N.~Saraf and A.~Bemporad, ``Fast model predictive control based on linear input/output models and bounded-variable least squares,'' in \emph{56th IEEE Conference on Decision and Control}.\hskip 1em plus 0.5em minus 0.4em\relax IEEE, 2017, pp. 1919--1924.

\bibitem{fletcher2000practical}
R.~Fletcher, \emph{Practical Methods of Optimization}.\hskip 1em plus 0.5em minus 0.4em\relax Chichester, England: John Wiley \& Sons, 2000.

\bibitem{boyd2004convex}
S.~P. Boyd and L.~Vandenberghe, \emph{Convex {O}ptimization}.\hskip 1em plus 0.5em minus 0.4em\relax Cambridge, England: Cambridge University Press, 2004.

\bibitem{max1950inverting}
A.~W. Max, ``Inverting modified matrices,'' in \emph{Memorandum Rept. 42, Statistical Research Group}.\hskip 1em plus 0.5em minus 0.4em\relax Princeton University, 1950, p.~4.

\bibitem{fine2001efficient}
S.~Fine and K.~Scheinberg, ``Efficient {SVM} training using low-rank kernel representations,'' \emph{Journal of Machine Learning Research}, vol.~2, pp. 243--264, 2001.

\bibitem{goldfarb2004product}
D.~Goldfarb and K.~Scheinberg, ``A product-form {C}holesky factorization method for handling dense columns in interior point methods for linear programming,'' \emph{Mathematical Programming}, vol.~99, no.~1, pp. 1--34, 2004.

\bibitem{goldfarb2005product}
------, ``Product-form {C}holesky factorization in interior point methods for second-order cone programming,'' \emph{Mathematical Programming}, vol. 103, no.~1, pp. 153--179, 2005.

\bibitem{goldfarb2008numerically}
------, ``Numerically stable {$LDL^\top$} factorizations in interior point methods for convex quadratic programming,'' \emph{IMA Journal of Numerical Analysis}, vol.~28, no.~4, pp. 806--826, 2008.

\bibitem{wright1997stability}
S.~Wright, ``Stability of augmented system factorizations in interior-point methods,'' \emph{SIAM Journal on Matrix Analysis and Applications}, vol.~18, no.~1, pp. 191--222, 1997.

\bibitem{wachter2006implementation}
A.~W{\"a}chter and L.~T. Biegler, ``On the implementation of an interior-point filter line-search algorithm for large-scale nonlinear programming,'' \emph{Mathematical programming}, vol. 106, pp. 25--57, 2006.

\bibitem{bemporad1997nonlinear}
A.~Bemporad, A.~Casavola, and E.~Mosca, ``Nonlinear control of constrained linear systems via predictive reference management,'' \emph{IEEE transactions on Automatic Control}, vol.~42, no.~3, pp. 340--349, 1997.

\bibitem{xie2017differential}
Z.~Xie, C.~K. Liu, and K.~Hauser, ``Differential dynamic programming with nonlinear constraints,'' in \emph{IEEE International Conference on Robotics and Automation}, 2017, pp. 695--702.

\bibitem{wang2010fast}
Y.~Wang and S.~Boyd, ``Fast evaluation of quadratic control-{L}yapunov policy,'' \emph{IEEE Transactions on Control Systems Technology}, vol.~19, no.~4, pp. 939--946, 2010.

\bibitem{ames2016control}
A.~D. Ames, X.~Xu, J.~W. Grizzle, and P.~Tabuada, ``Control barrier function based quadratic programs for safety critical systems,'' \emph{IEEE Transactions on Automatic Control}, vol.~62, no.~8, pp. 3861--3876, 2016.

\bibitem{xiao2021high}
W.~Xiao and C.~Belta, ``High-order control barrier functions,'' \emph{IEEE Transactions on Automatic Control}, vol.~67, no.~7, pp. 3655--3662, 2021.

\bibitem{bennett1965triangular}
J.~M. Bennett, ``Triangular factors of modified matrices,'' \emph{Numerische Mathematik}, vol.~7, pp. 217--221, 1965.

\bibitem{fletcher1974modification}
R.~Fletcher and M.~J. Powell, ``On the {M}odification of {$LDL^\top$} {F}actorizations,'' \emph{Mathematics of Computation}, vol.~28, no. 128, pp. 1067--1087, 1974.

\bibitem{gill1975methods}
P.~E. Gill, W.~Murray, and M.~A. Saunders, ``Methods for computing and modifying the {$LDV$} factors of a matrix,'' \emph{Mathematics of Computation}, vol.~29, no. 132, pp. 1051--1077, 1975.

\end{thebibliography}
\end{document}